\documentclass[preprint,showpacs,preprintnumbers,amsmath,amssymb]{revtex4}
\usepackage{graphicx}
\usepackage{subfigure}

\def\av#1{{\langle #1 \rangle}}
\newcommand{\be}{\begin{equation}}
\newcommand{\ee}{\end{equation}}
\newcommand{\F}{{\mathcal F}}
\newcommand{\Fo}{{\mathcal F}_o}

\newcommand{\calN}{\mathcal{N}}

\begin{document}

\title{Mock LISA data challenge for the Galactic white dwarf binaries}

\author{Arkadiusz B\l aut}
\affiliation{Institute of Theoretical Physics, University of Wroc\l
aw, Pl.Maxa Borna 9,
Pl-50-204  Wroc\l aw, Poland}

\author{Stanislav Babak}
\affiliation{Albert Einstein Institute, Golm, Am Muchlenberg 1,
D-14476 Golm bei Potsdam, Germany}

\author{Andrzej Kr\'olak}
\affiliation{Institute of Mathematics,
Polish Academy of Sciences,
\'Sniadeckich 8, 00-950 Warsaw, Poland \\
and \\
The Andrzej So\l tan Institute
for Nuclear Studies, 05-400 \'Swierk-Otwock, Poland}

\begin{abstract}
We present data analysis methods used in detection and the estimation of parameters of gravitational
wave signals from the white dwarf binaries in the mock LISA data challenge. Our main focus is on the
analysis of challenge 3.1, where the gravitational wave signals from more than $6 \times 10^7$ Galactic binaries
were added to the simulated Gaussian instrumental noise. Majority of the signals at low frequencies are not resolved
individually. The confusion between the signals is strongly reduced at frequencies above 5 mHz. Our basic
data analysis procedure is the maximum likelihood detection method. We filter the data through the template bank
at the first step of the search, then we refine parameters using the Nelder-Mead algorithm, we remove the strongest
signal found and we repeat the procedure. We detect reliably and estimate parameters accurately of more than
ten thousand signals from white dwarf binaries.
\end{abstract}


\maketitle

\section{Introduction}
\label{sec:intro}

The Galaxy contains a numerous population of the ultra-compact binaries which have orbital
periods shorter than one hour. The observations of those binaries carry important astrophysical
information about the internal stellar structure, formation of binaries and their evolution \cite{N09}.
Due to the short period and proximity, the ultra-compact Galactic binaries will be an important source
of gravitational waves (GW) for the future space borne interferometer LISA. LISA  is planned to be launched
in the next decade jointly by ESA and NASA. The bandwidth of the LISA detector is expected to be from 0.1 mHz to
100 mHz. In this frequency range we expect around $6 \times 10^7$ ultra-compact binaries.
These binaries will be dominated by the population of the white dwarf binaries. The number of the
observed ultra-compact binaries will be so large below 3 mHz that they are not individually resolvable
and form a cyclo-stationary background which dominates over the instrumental noise above 0.1 mHz
\cite{ETK05,ETK05b}. The number of binaries drops significantly above 7-8 mHz.

The  most common sources are  white-dwarf/white-dwarf binaries emitting gravitational wave signals
of nearly constant frequency and amplitude. We also expect to observe few white-dwarf/neutron star binaries.
The binaries could be of two major types:

(i) Detached, separated white-dwarf/white-dwarf binaries whose evolution is driven by radiation reaction. They are the end points of many binary evolution scenarios. The gravitational wave carry information about
the mass of the binary and the distance.

(ii) Interacting binaries. Those are close systems with a significant tidal interaction and/or with the Roche
lobe overflow. In those systems the gravitational radiation reaction competes against mass transfer and the
orbital period can either increase or decrease. Currently there are 22 known accreting binaries, so called
AM CVn stars, with periods between 5.4 and 65 min \cite{N09b}. The radiation from those binaries fall into
the LISA band and they will serve as {\it verification} binaries \cite{SV06}.

It was demonstrated \cite{rep:mldc2} that one can detect and remove a few tens of thousand of those signals.
The resolved systems will provide the map of the compact binaries in the Galaxy and will allow us to constrain
the evolutionary pathways of those systems.

A series of mock LISA data challenges (MLDC) was organized in order to foster development of LISA data
analysis algorithms and to compare the performance of different methods  \cite{web:mldc,over:mldc1,rep:mldc1,over:mldc2}. The simulated
Galaxy consists of only white dwarf binaries (of both types detached and interacting) in circular orbit
and it is based on the population synthesis described in \cite{N01,N04}.
It is expected that white dwarf binaries will largely dominate over more heavy binaries
like neutron stars and/or black holes. Eccentric white dwarf binaries could be generated in globular clusters
but their number is a tiny fraction of the total population \cite{B01,W07}.
The gravitational wave signals produced by white dwarf Galactic binaries span the whole LISA band, they stand above
the instrumental noise starting at 0.1 mHz and propagate all the way up to few tens of mHz.

Let us give a brief overview of currently available methods. The fully coherent methods employing
matched filtering techniques can be split in two groups: stochastic search and grid-based search.

The stochastic search does not map uniformly the whole parameter space, instead it concentrates on the regions
with high likelihood. The first type of the stochastic search is suggested in \cite{CCR07} and it is based on
the genetic algorithm. The genetic algorithm is an optimization method which evolves the set of templates
(a colony of organisms) in the direction of increasing likelihood (improving fitness of organisms) using a certain
rules (selection, breeding, mutation). Another stochastic method  is based on constructing the
(Markov) chains using Metropolis-Hastings acceptance/rejection rule. The Bayesian methods are powerful tools
to get posterior probability distribution function. A pure MCMC (Markov chain Monte-Carlo) algorithm does not perform
well due to presence of quite strong (and well separated in the parameter space) secondary maxima in the likelihood.
Two currently available Bayesian algorithms differ mainly in the way they explore the parameter space. BAM algorithm
suggested in \cite{C2} uses multiple proposal distributions reflecting possible correlations in the parameter space
in combination with the simulated annealing. At the search stage (the search for the global maximum in the
parameter space) the Markovian properties of the chain are quite often not respected, and, once the search
is completed, the sampling stage starts during which  the classical MCMC algorithm is used. The key feature of BAM is
blocking: the search is conducted in the several frequency bands each split in several small blocks. The algorithm
steps through these blocks updating all sources within a given band simultaneously. After all blocks have been
updated, they are shifted by one-half the width of a blocks for the next round of updates to eliminate
boundary effect. The results from the different frequency bands are glued together using the overlapped
buffer zones. Another MCMC-based method is described in \cite{TVV}. The authors suggest to use delayed
rejection MCMC to explore the parameter space, find the global maximum and sample the posterior distribution.
The basic idea behind "delayed rejection" is an extended acceptance/rejection rule: we allow several jumps
before rejecting/accepting the next point in the chain, and each following trial jump learns the property
of the parameter space explored by the previous jumps. An additional advantage of the "delayed rejection"
is that the variance of an estimate made from a chain using delayed rejection is always smaller than that
produced with a standard Markov chain.

The grid-based search maps the whole parameter space by computing the (log) likelihood (usually in the form of
$\mathcal{F}$-statistic)  on the uniformly distributed grid points (often referred as a template bank).
A particular implementation of this method is presented in this paper. Another version of the grid based
method is described in \cite{WPK}. There the authors adopt the software used for searching continuous GW
signals in the LIGO/VIRGO/GEO600 data to construct the grid in the parameter space (in four Doppler
parameters: sky location, frequency of GW signals and its derivative at some fiducial time) and to compute
$\mathcal{F}$- statistic.
Besides that, the authors in \cite{WPK} attempt to detect several signals at once while we are
dealing with one (the strongest) signal at the time.
They separate the secondary maxima from the primary by requiring the coincidence in the parameters recovered from
the analysis of different time delay interferometry (TDI) streams (for more information on TDI,
see \cite{TD} and references therein). Once the prime maxima are  identified, the grid mesh is refined
by zooming in onto them to improve the parameter estimation. The detected signals are removed from
the data and the procedure is repeated.

The last type of search uses Radon transform to  identify the frequency of the signal and the source's sky
location \cite{MN}. The basis of this method is that the LISA response function can be seen as a Radon
transform of binary distribution in those three parameters.

Galactic binaries were present in all four challenges conducted so far. In this article we report our
results of the analysis of challenge 3.1 data set.
This data set contains approximately $6\times 10^7$ Galactic binaries
with simulated instrumental noise. The GW signals had measurable frequency evolution at high frequencies.
The participants of the challenge have returned the parameters of the detected signals: the sky position
in ecliptic coordinates, frequency of GW and its first derivative, inclination of the orbit to the line of sight,
polarization angle, initial GW phase and the amplitude.

The paper is organized as follows. In Section 2 we shall present an analytic approximation to the response
of the LISA detector to a gravitational wave signal from a binary system. We shall not give details of
the derivation, these can be found in the original papers \cite{AET99}, \cite{CR03}, and \cite{KTV04}.
In Section 3 we shall present the maximum likelihood method in application to detection and estimation
of parameters of the GW signal from a binary system imbedded in stationary Gaussian noise.
In Section 4 we describe our data analysis tools and algorithms that we have implemented our computer
codes for challenge 3.1 search. We describe the search strategy in Section 5 and discuss the results
of the search in Section 6.

\section{Response of the LISA detector to the gravitational-wave signal from a binary system}
\label{sec:res} The LISA detector consists of 3 satellites forming a
constellation of an approximately equilateral triangle. The
constellation rotates around the Sun with a period of 1 year
trailing the Earth by 20 degrees. The triangular constellation is
inclined at 60 degrees to the ecliptic and rotates itself around
its center with a period of 1 year in the direction opposite to the
rotation around the Sun. The LISA detector in general will produce 3
independent data streams. In the long wavelength approximation, when
the length of the gravitational wave is much longer than the
distance between the spacecraft, the number of independent data
streams degenerates into 2. There are various combinations of the
responses of the LISA detector (see \cite{AET99} for details). The
data simulated for the mock LISA data challenge are the
first-generation TDI Michelson
combinations that we denote by {\it X}, {\it Y}, and {\it Z}. TDI is a software
technique (\cite{AET99}) by which we remove the dominant frequency
noise from the LISA instrumental noise, first generation means that
in the TDI procedure we assume that the distances between the
spacecraft are constant, independent of time. In this Section we
shall summarize approximate analytic formulas for the first TDI
generation Michelson responses of the LISA detector to a
gravitational wave signal from a binary system. The formulas are
essentially the same as in Appendix C \cite{KTV04} except that they
are given using the conventions used in the Synthetic LISA numerical
software \cite{V05}. The GW response of the first-generation TDI
Michelson observable $X$ is given by a linear combination of the
four time-dependent functions $X^{(k)}(t)$.
\begin{equation}
X(t) = 2 \, \omega L \, \sin(\omega L) \sum_{k=1}^{4} a^{(k)} X^{(k)}(t),
\label{X}
\end{equation}
where $\omega$ is the angular gravitational wave frequency and $L$ is the distance between
the spacecrafts of the LISA detector.

Before we give functions $X^{(k)}(t)$ let us introduce few
notations.  First we define the polarization basis in the solar
system barycenter frame (following \cite{how:mldc}):
\begin{eqnarray}
\hat{u} &=& \{ \sin(\beta),\cos(\lambda), \sin(\beta)\} \sim
\frac{\partial \hat{k}}{\partial \beta}\\
\hat{v} &=& \{ \sin(\lambda), -\cos(\lambda), 0 \} \sim
\frac{\partial \hat{k}}{\partial \lambda} ,
\end{eqnarray}
where  $\beta$ and $\lambda$ are, respectively, the latitude
and the longitude of the source in ecliptic coordinates
and $\hat{k} = -\{ \cos(\beta)\cos(\lambda),
\cos(\beta)\sin(\lambda), \sin(\beta)\}$ is direction of the wave
propagation. Then we introduce the LISA motion used in the
production of MLDC data. The position of each spacecraft can be
split in the position of the guiding center $\vec{R}$ and position
of the spacecraft with respect to that center:
\begin{equation}
\vec{r}_i = \vec{R} + L\vec{q}_i,\quad i=1,2,3,
\end{equation}
here we have assumed LISA to be a rigid equilateral triangle:
$L=L_1=L_2=L_3$, and the $\vec{q}_i$ are as follows

\begin{eqnarray}
\vec{q}_i =  \frac1{2\sqrt{12}}\left\{ \cos(2\Omega t  - \chi_i) -
3\cos(\chi_i), \sin(2\Omega t  - \chi_i) - 3\sin(\chi_i),
-\sqrt{12}\cos(\Omega t - \chi_i)  \right\},
\end{eqnarray}
where $\chi_i = 2(i-1)\pi/3$. The unit vectors
along the arms can be defined via vectors $\vec{q}_i$: $\hat{n}_1 =
\vec{q}_2 - \vec{q}_3$, and others are obtained by cyclic
permutation of indices:  $1\to 2\to 3 \to 1$. Now we are ready to
write $X^{(k)}$:

\begin{eqnarray}
\left[\!\begin{array}{c}
X^{(1)}  \nonumber \\
X^{(2)}
\end{array}\!\right]
= \left[\!\begin{array}{c}
u_2(t)   \nonumber \\
v_2(t)
\end{array}\!\right]\!
\bigl\{\mbox{sinc}\bigl[(1+\hat{k}\hat{n}_2)x/2\bigr]\cos\bigl[\phi(t) + (x/2)\hat{k}\vec{q}_2   - 3x/2\bigr] \\
\hspace{29mm}              + \hspace{1.5mm}
\mbox{sinc}\bigl[(1-\hat{k}\hat{n}_2)x/2\bigr]\cos\bigl[\phi(t) +
(x/2) \hat{k}\vec{q}_2    -
              5x/2\bigr]\bigr\} \nonumber
\\
\hspace{14mm} - \left[\!\begin{array}{c}
u_3(t) \nonumber \\
v_3(t)
\end{array}\!\right]\!
\bigl\{\mbox{sinc}\bigl[(1+\hat{k}\vec{n}_3)x/2\bigr]\cos\bigl[\phi(t) + (x/2)\hat{k}\vec{q}_3  - 5x/2\bigr] \\
\hspace{29mm}              + \hspace{1.5mm}
\mbox{sinc}\bigl[(1-\hat{k}\vec{n}_3)x/2\bigr]\cos\bigl[\phi(t) +
(x/2)\hat{k}\vec{q}_3  -
             3x/2\bigr]\bigr\}
\end{eqnarray}
and $X^{(3)}$, $X^{(4)}$ are obtained by replacing $\cos$ with $\sin$ in $X^{(1)}$ and $X^{(2)}$ respectively.
In the above equation we have used:
\begin{eqnarray}
u_i &=& -\frac1{2}\left[ (\hat{u} \hat{n}_i)^2 - (\hat{v} \hat{n}_i)^2\right]\\
v_i &=&  (\hat{u} \hat{n}_i) (\hat{v} \hat{n}_i),
\end{eqnarray}
where $x = \omega L$ and ${\mathrm{sinc}(\ldots)}\, =
\frac{\sin{(\ldots)}}{(\ldots)}$. The GW responses for $Y$ and $Z$
can be obtained by cyclical permutation of the spacecraft indices.

The phase modulation function $\phi(t)$ is given by
\begin{eqnarray}
\phi(t) = \omega t + \frac{1}{2}\dot{\omega} t^2 + (\omega  +
\dot{\omega} t)\, R \cos\beta\cos(\Omega t + \eta_o - \lambda),
\label{eq:dopplershifting}
\end{eqnarray}
where $\Omega = 2\pi/1$year, $\eta_o$ is the position of the constellation on the
orbit around the Sun at time $t = 0$, and $R$ is $1$ astronomical
unit. The small eccentricity of the Earth orbit (e = 0.017) can be neglected
because it contributes less than one cycle to the phase $\phi(t)$ above
for gravitational wave frequencies from 0.1 mHz to 12 mHz for which we
analyze the data. The parameter $\dot{\omega}$ is the frequency drift which may
occur either due to the gravitational radiation reaction or as a
result of the tidal interaction between the components of the binary
system. In the case of a detached binary system evolving only due to
the gravitational radiation reaction the frequency drift
$\dot{\omega}$ is approximately given by (see Section IID of
\cite{KTV04} for discussion).
\begin{equation}
\dot{\omega} = \frac{48}{5}\left(\frac{G {\cal M}_c}{2 c^3}\right)^{5/3}\!\!\!\omega^{11/3},
\end{equation}
where $\mathcal{M}_c = m_1^{3/5} m_2^{3/5} / (m_1 + m_2)^{1/5}$
is the chirp mass ($m_1$ and $m_2$ are the individual  masses of the components of the binary).

Finally the constant amplitudes $a^{(k)}$ take the form
\begin{eqnarray}
a^{(1)} &=&  \ h_0^+\cos\phi_0 \,\cos 2\psi -  h_0^\times\sin\phi_0 \,\sin 2\psi, \label{eq:ampone} \\
a^{(2)} &=&  \ h_0^+\cos\phi_0 \,\sin 2\psi +  h_0^\times\sin\phi_0 \,\cos 2\psi, \\
a^{(3)} &= - & \ h_0^+\sin\phi_0\,\cos 2\psi  -  h_0^\times\cos\phi_0\,\sin 2\psi, \\
a^{(4)} &= - & \ h_0^+\sin\phi_0\,\sin 2\psi  +
h_0^\times\cos\phi_0\,\cos 2\psi, \label{eq:ampfour}
\end{eqnarray}
where
\begin{eqnarray}
\label{eq:hpc}
h_0^+ &=& h_0(1 + \cos^2\iota)/2, \\
h_0^\times &=& h_0\cos\iota.
\end{eqnarray}
The parameters $h_0$, $\phi_0$, $\psi$, and $\iota$ are constant
amplitude, the constant phase of the signal, the polarization angle,
and the inclination angle respectively. In the case of a detached
binary system evolving only due to the gravitational radiation
reaction the constant amplitude $h_0$ is given by
\begin{equation}
h_0 = \frac{4 (G \mathcal{M}_c)^{5/3}}{c^4 D_L} \left[ \frac{\omega}{2} \right]^{2/3} \! \!,
\end{equation}
where $D_L$ is the luminosity distance to the source.

One can invert the equations (\ref{eq:ampone}) for amplitudes to obtain formulas
for astrophysical parameters $h_0$, $\phi_0$, $\psi$, and $\iota$.
We first introduce the quantities
\begin{eqnarray}
A &=& (a^{(1)})^2 + (a^{(2)})^2 + (a^{(3)})^2 + (a^{(4)})^2, \\
D &=& a^{(1)} a^{(4)} - a^{(2)} a^{(3)}.
\end{eqnarray}
Then the constants $h_0^+$, $h_0^+$, $h_0$, and $\phi_0$
can be uniquely determined.
\begin{eqnarray}
h_0^+ &=& \sqrt{(A + \sqrt{A^2 - 4D^2})/2}, \\
h_0^\times &=& \mbox{sign}(D)\sqrt{(A - \sqrt{A^2 - 4D^2})/2}, \\
h_0 &=& h_0^{+} + \sqrt{h_0^{+2} - h_0^{\times 2}}, \\
\iota &=& \mbox{acos}(h_0^\times/h_0).
\end{eqnarray}
Finally the constant phase $\phi_0$ and the polarization angle $\psi$ can be obtained
from the following equations:
\begin{eqnarray}
\tan2\phi_0 &=& \frac{2(a^{(1)} a^{(3)} + a^{(2)}a^{(4)})}
           {(a^{(3)})^2+(a^{(4)})^2 - (a^{(1)})^2 - (a^{(2)})^2},
\\[1ex]
\tan4\psi &=& \frac{2(a^{(1)} a^{(2)} + a^{(3)}a^{(4)})}
       {(a^{(1)})^2+(a^{(3)})^2 - (a^{(2)})^2 - (a^{(4)})^2}.
\end{eqnarray}

The long-wavelength ({\rm LW}) approximation to the GW responses is
obtained by taking the leading-order terms of the generic
expressions in the limit of $\omega L \rightarrow 0$:
\begin{eqnarray}
X_{LW}(t) & \simeq & 4(\omega L)^2 \{[u_2(t)-u_3(t)] \left[ a^{(1)}
\cos \phi(t) + a^{(3)} \sin \phi(t) \right] \\
&& +[v_2(t)-v_3(t)] \left[ a^{(2)} \cos \phi(t) + a^{(4)} \sin
\phi(t) \right]\} \label{eq:xlw}
\end{eqnarray}
and $Y_{LW}$, $Z_{LW}$ responses are obtained by cyclical
permutation of the indices.

In Fig. 1 we have compared the power spectra of the signal
generated using the analytic formulas given above by equation \eqref{X} with the power
spectrum of noise free response generated by Synthetic LISA software
for one of the training sets provided in MLDC. Synthetic LISA
software was used to generate the analyzed MLDC data sets.

\begin{figure}[t]
\centering
\includegraphics{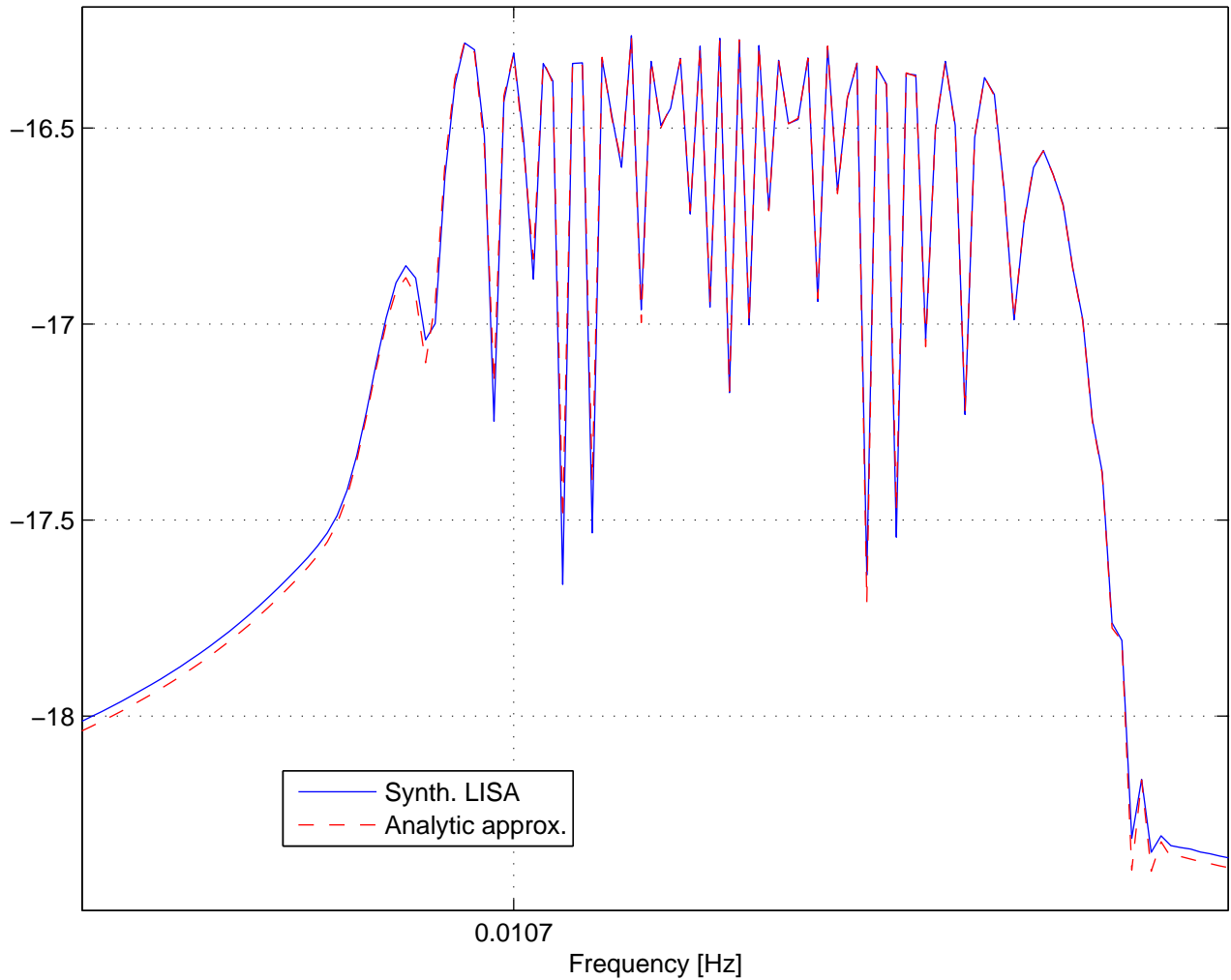}
\caption{Comparison of the power spectra of the gravitational-wave signal
response using the analytic formulas presented in this paper and {\em
Synthetic LISA} \cite{V05}.}
\end{figure}

\section{Maximum likelihood detection and parameter estimation}
\label{sec:filter}
To detect the signal and estimate its parameters we use the {\em maximum likelihood
(ML) estimation} method which consists of maximizing the likelihood function $\Lambda$
with respect to the parameters of the signal \cite{W71}. Let us first consider the Michelson $X$ combination
given in the previous section. We assume that the noise in the detector is a stationary Gaussian random process.
Moreover we assume that over the bandwidth of the signal the spectral density of the noise
is approximately constant and equal to $S_o = S(\omega_o)$. This condition should be well fulfilled
for the case of a gravitational wave signal from a white dwarf binary in LISA detector noise. Then we can
approximate the log likelihood function by
\begin{equation}
\label{eq:LFpul}
\log\Lambda = 2\,\frac{T_o}{S_o} [ \av{y^X X}  - \frac{1}{2} \av{X^2} ],
\end{equation}
where $y^X$, is the noisy data in the $X$ channel, $S_o$ is
one-sided spectral density of the noise, $T_o$ is the observation
time, and the time-averaging operator $\av{\cdot}$ is defined by
\begin{equation}
\av{g} := \frac{1}{T_o}\int^{T_o}_{0}g(t)\,dt.
\end{equation}
By introducing the variables
\begin{equation}
h^{(k)} = 2 \, \omega L \, \sin(\omega L) X^{(k)}
\end{equation}
the $X$ combination can be written in a compact form
\begin{equation}
\label{eq:Xah1}
X = \sum_{k=1}^4 a^{(k)} h^{(k)}.
\end{equation}
The ML estimators $\hat{a}^{(k)}$ of the amplitudes are found by
maximizing $\log \Lambda$ with respect to parameters $a^{(k)}$, that
is by solving
\begin{equation}
\label{eq:MLa}
\frac{\partial\log\Lambda}{\partial a^{(k)}} = 0.
\end{equation}
The equations (\ref{eq:MLa}) above are equivalent to the following set of
linear equations
\begin{equation}
\sum_{k=1}^4 M^{(l)(k)} a^{(k)} = N^{(l)},\quad l=1,\ldots,4,
\label{eq:linsys}
\end{equation}
where
\begin{eqnarray}
\label{eq:mn1}
M^{(l)(k)} &=& \av{h^{(k)}\,h^{(l)}}, \label{eq:mdef} \\
N^{(l)} &=& \av{y^X\,h^{(l)}}.
\label{eq:ndef}
\end{eqnarray}
Thus the maximum likelihood estimators $\hat{a}^{(k)}$ of the amplitudes are explicitly given by
\begin{equation}
\hat{a}^{(k)} = \sum_{l=1}^{4}\big(M^{-1}\big)^{(l)(k)} N^{(l)}.
\end{equation}
Substituting the above estimators $\hat{a}^{(k)}$ for amplitudes $a^{(k)}$
in the log likelihood function log$\Lambda$ yields the reduced log likelihood function
that we denote by $\mathcal{F}$:
\begin{equation}
\mathcal{F} = {\textstyle \frac{T_o}{S_o}} \sum_{l=1}^{4} \sum_{k=1}^{4}
\big(M^{-1}\big)^{(l)(k)} N^{(l)} N^{(k)}.
\end{equation}
We call the above function the  $\mathcal{F}$-statistic.

One can show that the following relations hold approximately for the components
of the matrix $M^{(l)(k)}$
\begin{eqnarray}
\label{eq:app0}
\av{h^{(1)}\,h^{(3)}} &=&  \av{h^{(2)}\,h^{(4)}} = 0, \\
\label{eq:Ua}
\av{h^{(1)}\,h^{(1)}} &=&  \av{h^{(3)}\,h^{(3)}}, \\
\label{eq:Va}
\av{h^{(2)}\,h^{(2)}} &=&  \av{h^{(4)}\,h^{(4)}}, \\
\label{eq:Qa}
\av{h^{(1)}\,h^{(2)}} &=&  \av{h^{(3)}\,h^{(4)}}, \\
\label{eq:Pa}
\av{h^{(1)}\,h^{(4)}} &=& -\av{h^{(2)}\,h^{(3)}}.
\end{eqnarray}
It is convenient to introduce the following variables
\begin{eqnarray}
\label{eq:U}
U =  2 \av{h^{(1)}\,h^{(1)}}, \\
\label{eq:V}
V =  2 \av{h^{(2)}\,h^{(2)}}, \\
\label{eq:Q}
Q =  2 \av{h^{(1)}\,h^{(2)}}, \\
\label{eq:P}
P =  2 \av{h^{(1)}\,h^{(4)}}.
\end{eqnarray}

Let us introduce the complex amplitude parameters
\begin{eqnarray}
\label{eq:CA}
a^{(u)} &=& a^{(1)} + i a^{(3)}, \\
\label{eq:CB}
a^{(v)} &=& a^{(2)} + i a^{(4)},
\end{eqnarray}
where $a^{(k)}$ are given by
Eqs.\,(\ref{eq:ampone})--(\ref{eq:ampfour}). Let us also define the
complex modulation functions $m^{(u)}$ and $m^{(v)}$ by
\begin{equation}
\label{eq:mumv}
\begin{split}
\left[\!\begin{array}{c}
m^{(u)} \\
m^{(v)}
\end{array}\!\right]
=&
\left[\!\begin{array}{c}
u_2(t) \\
v_2(t)
\end{array}\!\right]\!
\bigl\{\mbox{sinc}\bigl[(1+\hat{k}\vec{n}_2)x/2\bigr]\exp i \bigl[(x/2) \hat{k}\vec{q}_2 - 3x/2\bigr] \\
& \quad\quad\quad\quad + \mbox{sinc}\bigl[(1-\hat{k}\vec{n}_2)x/2\bigr]
\exp i \bigl[(x/2) \hat{k}\vec{q}_2 - 5x/2\bigr]\bigr\} -\\
& \left[\!\begin{array}{c}
u_3(t) \\
v_3(t)
\end{array}\!\right]\!
\bigl\{\mbox{sinc}\bigl[(1++\hat{k}\vec{n}_3)x/2\bigr]\exp i \bigl[(x/2) \hat{k}\vec{q}_3 - 5x/2\bigr]\\
& \quad\quad\quad\quad +
\mbox{sinc}\bigl[(1-\hat{k}\vec{n}_3)x/2\bigr]\exp i \bigl[(x/2)
\hat{k}\vec{q}_3 - 3x/2\bigr]\bigr\}.
\end{split}
\end{equation}
Introducing further complex quantities
\begin{eqnarray}
\label{eq:CQ}
W   &=& Q + i P, \\
\label{eq:CFa}
N^{(u)} &=& N^{(1)} + i N^{(3)}, \\
\label{eq:CFb}
N^{(v)} &=& N^{(2)} + i N^{(4)},
\end{eqnarray}
where $N^{(k)}$, $Q$ and $P$ are given by
Eqs.\,(\ref{eq:ndef}), (\ref{eq:Q}), and (\ref{eq:P}) respectively
and using the approximate relations given by Eqs.\,
(\ref{eq:app0}), (\ref{eq:Ua}), (\ref{eq:Va}), (\ref{eq:Qa}), and (\ref{eq:Pa})) we
can write the ML amplitude estimators of the complex amplitudes and
the $\F$-statistic in the following compact form
\begin{equation}
\label{amlec}
\left(\begin{array}{c}
\hat{a}^{(u)} \\ \hat{a}^{(v)}
\end{array}\right) = \frac{2}{\Delta}
\left(\begin{array}{cc}
 V & -W^*  \\
-W &  U
\end{array}\right) \cdot
\left(\begin{array}{c}
N^{(u)} \\ N^{(v)}
\end{array}\right),
\end{equation}
\begin{equation}
\label{eq:mld3} {\mathcal F} = 2 \frac{T_o}{S_o} \frac{\left\{ V
\bigl|N^{(u)}\bigr|^2 + U \bigl|N^{(v)}\bigr|^2
               - 2 \, \mathrm{Re} \left[ W \, N^{(u)} (N^{(v)})^* \right] \right\}}{\Delta}.
\end{equation}
where $\Delta = U V - |W|^2$.
The integrals $N^{(u)}$ and $N^{(v)}$ can be expressed as
\begin{eqnarray}
N^{(u)} &=& 2 \, \omega L \, \sin(\omega L) \av{y^X(t) \, m^{(u)}(t) \,
\exp i \phi(t)} , \\
\label{eq:mu}
N^{(v)} &=& 2 \, \omega L \, \sin(\omega L) \av{y^X(t) \, m^{(v)}(t) \,
\exp i \phi(t)}.
\label{eq:mv}
\end{eqnarray}
As we shall see in the next section the above form of the  $\mathcal{F}$-statistic
is very suitable for a numerical implementation. In a similar manner one can derive
the ${\mathcal F}$-statistic for other Michelson variables.

In order to extract information about the signal from all three
independent variables we need to derive the $\F$-statistic
for the whole LISA network. It is then useful to consider
the so called "optimal" combinations of the responses. These
combinations have the property that their instrumental noises are
uncorrelated (see \cite{P02}) and consequently their cross-spectrum
matrix is diagonal. In this case the log likelihood function for the
whole network is the sum of log likelihood functions for the
individual combinations,
For the Michelson variables the optimal combination are given by
\cite{optimal}
\begin{eqnarray}
A &=& \frac{Z - X}{\sqrt{2}}, \\
E &=& \frac{X - 2Y + Z}{\sqrt{6}}, \\
T &=& \frac{X + Y + Z}{\sqrt{3}}.
\end{eqnarray}
The noisy data, $y^A,y^E$ and $y^T$ are obtained as the analogous combination of
the Michelson observables $y^X,y^Y$ and $y^Z$, where $y^Y$, $y^Z$ are data in the $Y$
and $Z$ channels respectively.
The log likelihood function for the network takes the form
\begin{equation}
\label{eq:vec1LFpul}
\begin{split}
\log{\cal L}  = 2\,T_o & \left\{\frac{1}{S_A(\omega_o)} [ \av{y^A A}  -
\frac{1}{2} \av{A^2}]+\right.\\
& \frac{1}{S_E(\omega_o)} [ \av{y^E E}  -
\frac{1}{2} \av{E^2}]+
\left.\frac{1}{S_T(\omega_o)} [ \av{y^T T}  -
\frac{1}{2} \av{T^2} ]\right\}.
\end{split}
\end{equation}
The power spectral densities $S_A, S_E, S_T$ are given by
\begin{equation}
\begin{array}{rcl}
S_A(\omega) & = & S_E(\omega)=32 \cos{(\omega L/2)}^2 \sin{(\omega L/2)}^2\left\{
[6+4\cos{(\omega L)}+\right.\\
&& \left.2\cos{(2\omega L)}]S^{pm}+[2+\cos{(\omega L)}]S^{op}\right\}\\
S_T(\omega) & = & 128\cos{(\omega L/2)}^2\sin{(\omega L/2)}^4\left[4\sin{(\omega L/2)}^2S^{pm}+S^{op}\right],
\end{array}
\end{equation}
where $S^{\rm pm}$ and $S^{\rm op}$ are spectral densities of proof-mass noise and optical path noise
respectively.
It turns out that the ML estimators of the amplitudes and the $\F$-statistic can be written
in the same form as for a single detector. To do this we introduce the following
noise-weighted average procedure. For any two vectorial quantities ${\bf p}$ and ${\bf q}$,
\begin{equation}
\begin{array}{rcl}
{\bf p}(t) & = &
\left(p^{A}(t), p^{E}(t), p^{T}(t)\right),\\
\quad
{\bf q}(t) & = &
\left(q^{A}(t), q^{E}(t), q^{T}(t)\right),
\end{array}
\end{equation}
the noise-weighted average operator $\av{\cdot}_{\cal S}$
is defined as follows,
\be
\av{{\bf p}\,{\bf q}}_{\cal S} := w_{A} \, \av{p^A\,q^A}
+ w_{E} \, \av{p^E\,q^E} + w_{T} \, \av{p^T\,q^T},
\ee
where the weights $w_I$ ($I=A,E,T$) are defined by
\begin{equation}
w_I := \frac{S^{-1}_I}{{\cal S}^{-1}}, \quad I=A,E,T, \quad
\text{with} \quad {\cal S}^{-1} = S_{A}^{-1} +  S_{E}^{-1} + S_{T}^{-1}.
\end{equation}
With the above definitions the log likelihood function of Eq.\,(\ref{eq:vec1LFpul})
defined for data \\ ${\bf y}=(y^A,y^E,y^T)$ and response ${\bf R}=(A,E,T)$
can be written in a compact form
\begin{equation}
\label{eq:vec2LFpul}
\log{\cal L} = 2\,\frac{T_o}{{\cal S}(\omega_o)} [ \av{{\bf y} {\bf R}}_{\cal S}  -
\frac{1}{2} \av{{\bf R}^2}_{\cal S} ].
\end{equation}
Having now defined ${\bf h}^{(k)}$ by [see Eq.\,(\ref{eq:Xah1})]
\begin{equation}
{\bf R} = \sum_{k=1}^4 a^{(k)} {\bf h}^{(k)}
\end{equation}
and ${\cal M}^{(k)(l)},\;{\cal N}^{(l)}$ by [see Eqs.\,(\ref{eq:mdef}),(\ref{eq:ndef})]
\begin{eqnarray}
\label{eq:mnvec1}
{\cal M}^{(k)(l)} &=& \av{{\bf h}^{(k)}\,{\bf h}^{(l)}}_{\cal S}, \\
{\cal N}^{(l)} &=& \av{{\bf y}\,{\bf h}^{(l)}}_{\cal S}.
\label{eq:mnvec2}
\end{eqnarray}
it is straightforward to get the maximum likelihood estimators of the
complex amplitudes and the $\F$-statistic for the LISA network
\begin{equation}
\label{amlecN}
\left(\begin{array}{c}
\hat{a}^{(u)}_{opt} \\ \hat{a}^{(v)}_{opt}
\end{array}\right) = \frac{2}{{\sf \Delta}}
\left(\begin{array}{cc}
 {\sf V} & -{\sf W}^*  \\
-{\sf W} &  {\sf U}
\end{array}\right) \cdot
\left(\begin{array}{c}
\calN^{(u)} \\ \calN^{(v)}
\end{array}\right),
\end{equation}
\begin{equation}
\label{eq:mld3N}
{\mathcal F}_{opt} =
2 \frac{T_o}{\cal S} \frac{\left\{ {\sf V} \bigl|\calN^{(u)}\bigr|^2 +
{\sf U} \bigl|\calN^{(v)}\bigr|^2
- 2 \, \mathrm{Re} \left[ {\sf W} \, \calN^{(u)} (\calN^{(v)})^* \right] \right\}}{{\sf \Delta}},
\end{equation}
where ${\sf \Delta} = {\sf U}{\sf V} - |{\sf W}|^2$ and $\sf W = \sf Q + i\sf P$.
The quantities ${\sf V}$, ${\sf U}$, ${\sf Q}$, $\sf P$ in Eqs.\,(\ref{amlecN}) and (\ref{eq:mld3N})
above are defined in the same way as quantities $V, U, Q, P$ in Eqs.\,(\ref{eq:U})--(\ref{eq:P}),
but with the time-averaging operator $\av{\cdot}$ replaced everywhere by noise-weighted
averaging operator $\av{\cdot}_S$ and scalar functions $h^{(k)}$ by their vectorial counterparts ${\bf h}^{(k)}$.
The two functions ${\cal N}^{(u)}$ and ${\cal N}^{(v)}$ are explicitly given by
$$
\begin{array}{rcll}
{\cal N}^{(u)} &=&
{\cal N}^{(1)}+i{\cal N}^{(3)}&=
2 \, \omega L \, \sin(\omega L) \av{{\bf y}(t) \, {\bf m}^{(u)}(t) \,
\exp i \phi(t)}_{\cal S} , \\
\label{eq:mu2}
{\cal N}^{(v)} &=&
{\cal N}^{(2)}+i{\cal N}^{(4)}&=
2 \, \omega L \, \sin(\omega L) \av{{\bf y}(t) \, {\bf m}^{(v)}(t) \,
\exp i \phi(t)}_{\cal S},
\label{eq:mv2}
\end{array}
$$
where two vector functions ${\bf m}^{(u)}$ and ${\bf m}^{(v)}$ are the
relevant combinations of $m^{(u)}$'s and $m^{(v)}$'s defined for $X$, $Y$ and $Z$
[see Eq.\,(\ref{eq:mumv})].
\section{Data analysis algorithms}
\label{sec:search}
\subsection{Use of the FFT algorithm}
\label{subsec:FFT}
The detection statistic, $\mathcal{F}$ [Eq.\ (\ref{eq:mld3}) or (\ref{eq:mld3N})], involves integrals
$N^{(u)}$ and $N^{(v)}$ of the form
\begin{equation}
{\cal I} = \int^{T_0}_0 y(t) \, m(t;\omega,\beta,\lambda) \,
\exp [i \phi_\mathrm{mod} (t;\omega,\dot{\omega},\beta,\lambda)]
\exp [i \omega t] \, dt
\label{fstat}
\end{equation}
where $m$ is one of the two complex modulation functions (\ref{eq:mumv}),
while the phase modulation $\phi_\mathrm{mod}$ is given by
\begin{equation}
\label{eqs:MNA}
\phi_\mathrm{mod}(t;\omega,\beta,\lambda) = \frac{1}{2}\dot{\omega} t^2 +
\omega R\cos\beta\cos(\Omega t + \eta_0 - \lambda)
\end{equation}
[see Eq.\ (\ref{eq:dopplershifting})]. In order to evaluate the integral (\ref{fstat}) efficiently
we would like to use the FFT algorithm. The integral (\ref{fstat}) is not a Fourier transform
because both the phase modulation function $\phi_\mathrm{mod}$ and the amplitude modulation function $m$
depend on the angular frequency $\omega$. We overcome these problems in the following way. Firstly we
introduce a new representation of the phase, namely we introduce two new parameters
\begin{eqnarray}
A   = \omega R \cos\beta\cos(\lambda - \eta_0), \nonumber\\
B   = \omega R \cos\beta\sin(\lambda - \eta_0).
\label{eq:AB}
\end{eqnarray}
In this new parametrization the phase modulation takes the form
\begin{equation}
\phi_\mathrm{mod}(t;\dot{\omega}, A, B) = \frac{1}{2}\dot{\omega} t^2 + A \cos(\Omega t) + B \sin(\Omega t),
\end{equation}
and it is independent of the angular frequency parameter $\omega$. A similar parametrization
has been used in the search of the resonant bar NAUTILUS detector detector data
for gravitational waves from spinning neutron stars (see \cite{ABJK02}). (As we
shall see in the following section the above parametrization by the
coordinates $\{\omega,\dot\omega,A,B\}$ will also prove useful in
the construction of our grid.) Secondly we assume that the bandwidth $[\omega_1 \, \omega_2]$
of the data is small so that the amplitude modulation function $m$ varies little over the interval
of angular frequency from lower edge of the band $\omega_1$ to the upper edge of the band $\omega_2$.
In order to  satisfy the above approximation, in our search
we have divided the data into narrow bands of
$\Delta f \equiv (\omega_2 - \omega_1)/(2\pi) = 0.1$ mHz by passing them
through narrowband filters (see Section \ref{sSearch} for details)
and each narrow-banded data were analyzed separately.
Then we can approximate the modulation function  $m(t;\omega,\beta,\lambda)$ by
$m(t;\omega_\mathrm{mid},A,B) =
m(t;\omega_\mathrm{mid},\beta_\mathrm{mid}(A,B,\omega_\mathrm{mid}),
\lambda_\mathrm{mid}(A,B,\omega_\mathrm{mid}))$ where
$\omega_\mathrm{mid} = (\omega_2 - \omega_1)/2$
and where $\beta_\mathrm{mid}(A,B,\omega_\mathrm{mid})$ and
$\lambda_\mathrm{mid}(A,B,\omega_\mathrm{mid})$ are obtained by
inverting Eqs.\,(\ref{eq:AB}) for $\omega = \omega_\mathrm{mid}$.
They are explicitly given by
\begin{eqnarray}
\label{eq:bemid} \beta_\mathrm{mid}(A,B,\omega_\mathrm{mid}) &=&
\pm \arccos\left(\frac{\sqrt{A^2 + B^2}}{\omega_\mathrm{mid} R}\right) , \\
\label{eq:lamid} \lambda_\mathrm{mid}(A,B,\omega_\mathrm{mid}) &=&
\eta_0 + \arctan\left(\frac{B}{A}\right).
\end{eqnarray}
Note that there are two values of the ecliptic latitude $\beta$ for
each pair of the values $A$ and $B$. Consequently the integral
(\ref{fstat}) can be approximated by
\begin{equation}
{\cal I} \simeq \int^{T_0}_0 y(t) \, m(t;\omega_\mathrm{mid},A,B) \,
\exp [i \phi_\mathrm{mod} (t;\dot\omega,A,B)] \exp [i \omega t] \,
dt, \label{fstata}
\end{equation}
For discrete data $y(t)$ the above integral can be converted to a discrete Fourier transform
which can be calculated by the FFT algorithm.

\subsection{Metric on the intrinsic parameters space}
In our method the detection of weak, quasi-monochromatic GW signals relies on an
efficient placement of the templates in the bank. It should minimize
the number of templates for a certain accepted  loss of a signal-to-noise ratio. Here we follow the
geometric approach initialized in \cite{Bala,Owen} and introduce a metric
on the intrinsic parameters space to measure the mismatch between the true
signal and the template.
In order to construct the metric and a grid on the intrinsic parameter space
over which we calculate the ${\cal F}$-statistic
we introduce an approximation to the signal that we call the
{\em linear model} \cite{AAS}:
\begin{equation}
\label{sm}
s(t)= A_0\cos{(\omega
t+\frac12\dot{\omega}t^2+A\cos{\Omega t}+B\sin{\Omega t}+\phi_0)},
\end{equation}
where $A_0$ is a constant amplitude and $\phi_0$ is a constant phase
and where $A$ and $B$ are parameters given by Eqs.\,(\ref{eq:AB}).
The phase modulation of the linear model is exactly the same as that of the
exact model whereas the amplitude is constant. This is a reasonable
approximation because the amplitude modulation functions vary very slowly, they
are periodic functions of one year.

The metric on intrinsic parameter space is defined by the reduced Fisher matrix
which is obtained from the full Fisher matrix of the linear model by projecting
on the intrinsic parameters space and normalizing it.
The reduced Fisher matrix $\tilde{\Gamma}$ determines the loss of
signal-to-noise ratio when parameters of the signal,
$\vec{\theta}=(\omega,\dot{\omega}, A, B)$, differ from the
parameters of the template by
$\Delta\vec{\theta}=(\Delta\omega,\Delta\dot{\omega},\Delta A,\Delta
B)$ \cite{JK,Prix}:
\begin{equation}
\label{eqn:Delpam}
\tilde{\Gamma}_{\vec{\theta}}(\Delta\vec{\theta},\Delta\vec{\theta})
= r^2 = \frac{\rho_{\vec{\theta}}^2(0)-
\rho_{\vec{\theta}}^2(\Delta\vec{\theta})}{\rho_{\vec{\theta}}^2(0)}\;\;+\;\;{\mbox
\O}(|\Delta\vec{\theta}|^3),
\end{equation}
where $\rho_{\vec{\theta}}(0)$ is the optimal signal-to-noise ratio
and $\rho_{\vec{\theta}}(\Delta\vec{\theta})$ is the signal-to-noise
ratio for the mismatch $\Delta\vec{\theta}$. The quantity $r^2$ is usually called
the {\em mismatch} and it is denoted by $m$ in \cite{Prix}.
For the calculations of the reduced Fisher matrix we refer the reader to Appendix \ref{app:lm}.
We show there that the approximation of the linear model leads to a particularly
simple forms of the reduced Fisher matrices on 3-dimensional intrinsic parameter
space $\{\omega,A,B\}$ (\ref{appFM3red}),
\begin{equation}\label{FM3red}
\tilde{\Gamma}_3 =
\left(%
\begin{array}{cccc}
  \frac{1}{12}              & 0                     & -\frac{1}{2 \pi n}   \\
  0                         & \frac12               & 0   \\
 -\frac{1}{2\pi n}          & 0                     & \frac12 \\
\end{array}%
\right)
\end{equation}
and 4-dimensional intrinsic parameter space $\{\omega,\dot\omega,A,B\}$ (\ref{appFM4red}),
\begin{equation}\label{FM4red}
\tilde{\Gamma} =
\left(%
\begin{array}{cccc}
  \frac{1}{12}              & \frac{1}{24}          & 0                     & -\frac{1}{2 \pi n}   \\
  \frac{1}{24}              & \frac{1}{45}          & \frac{1}{4\pi^2 n^2}  & -\frac{1}{4 \pi n} \\
  0                 &  \frac{1}{4\pi^2 n^2}         & \frac12 & 0   \\
 -\frac{1}{2\pi n}  & -\frac{1}{4 \pi n}            & 0   & \frac12 \\
\end{array}%
\right)  .
\end{equation}

We will use the reduced Fisher matrices (\ref{FM3red}) and (\ref{FM4red})
to build the grid in such a way that the distance defined
by $\tilde{\Gamma}_3$ or $\tilde{\Gamma}$ from any
point of the parameter space to the nearest node of the grid is not larger than some fixed value $r$.
We also see that not only the metrics $\tilde{\Gamma}_3$ and $\tilde{\Gamma}$ are flat but in the
coordinates  $\{\omega,A,B\}$ and $\{\omega,\dot\omega,A,B\}$ their coefficients are
constant, independent of the values of the parameters.

\subsection{Construction of the grid in the parameter space}
\label{subsec:grid}
\subsubsection{Covering problem on lattices}
\label{sCovering} The problem of constructing a grid in a
$d$-dimensional parameter space is equivalent to the problem of
covering $d$-dimensional space with equal overlapping spheres of a
given radius. The optimal covering would have minimal possible {\it
thickness} or {\it density} of covering defined as the average
number of spheres that contain a point of the space (see \cite{Con}
and below). When the metric is flat, as in the linear model, centers
of the spheres can lie on a $d$-dimensional {\it lattice}. In this
case one can take advantage of the theory of lattice coverings. In
the rest of the chapter we briefly sketch the basic definitions from
the theory of lattices that will be used in the construction of the
grid.

In general for any discrete set of points
${\cal S}=\left\{\vec{s}_1,\vec{s}_2,\ldots \right\}$ in ${\mathbb R}^n$
the {\it covering radius} $R$ of ${\cal S}$ is defined as the least
upper bound for any point of ${\mathbb R}^n$ to the closest point
$\vec{s}_i$:
$$
R({\cal S})=\sup_{\vec{x}\in{\mathbb R}^n}\inf_{\vec{s}\in{\cal
S}}|\vec{x}-\vec{s}|.
$$
Then spheres of equal radius $r$ centered at the points $\vec{s}_i$
will cover ${\mathbb R}^n$ only if $r\geq R$.

A {\it lattice} $\Lambda$ is a discrete subset of ${\mathbb R}^n$. Any
lattice has a {\it basis} $b=\left\{ \vec{b}_1, . . . , \vec{b}_n
\right\}$ of linearly independent vectors on ${\mathbb R}^n$ such that the lattice is
the set of all linear combinations of $\vec{b}_i$'s with integer
coefficients:
\begin{equation}
\label{lat1}
\Lambda=\left\{\sum_{i=1}^{n}c_i\vec{b}_i:\;\;\;c_i\in\mathbb{
Z},\;\;\;\;\;\; i=1,2,\ldots,n \right\}.
\end{equation}
A lattice basis is not unique, in dimensions $d>1$ there are
infinitely many of them, but all the bases have the same number of
elements called the {\it dimension} of the lattice.
To specify a basis $b$ of a lattice we will use the notation $\Lambda(b)$.

A lattice $\Lambda_1$ is {\it equivalent} to a lattice $\Lambda_2$ if
$\Lambda_1$ can be transformed into $\Lambda_2$ by a rotation
reflection and change of scale.

The parallelotope consisting of points
$c_1\vec{b}_1+\ldots+c_n\vec{b}_1$ with $0\leq c_i<1$ is a {\it
fundamental parallelotope} and is an example of an {\it elementary
cell}, that is the building block containing one lattice point which
tiles the whole ${\mathbb R}^n$ by translations of lattice vectors.
There are infinitely many elementary cells but the volume of each
elementary cell is unique for a given lattice $\Lambda$.

The {\it Voronoi cell} around any point $\vec{v}$ of $\Lambda$ is
the set of vectors $\vec{x}$ of ${\mathbb R}^n$ which are closer to
$\vec{v}$ than to any other lattice vector:
\begin{equation}
\label{vor1} V(\vec{v})=\left\{
\vec{x}:\;\;\;|\vec{x}-\vec{w}|\geq|\vec{x}-\vec{v}|\;\;\;
{\mbox{for all}}\;\;\;\vec{w}\in \Lambda  \right\}.
\end{equation}
All Voronoi cells of a given lattice are congruent convex polytopes
and are another examples of elementary cells sometimes referred to
as Wigner-Seitz cells or Brillouin zones.

For the lattice $\Lambda$ having Voronoi cells congruent to polytope
$V(\vec{v})$, where $\vec{v}$ is any of the lattice points, the
covering radius $R(\Lambda)$ is the circumradius of $V(\vec{v})$  \,
i.e. the largest distance between $\vec{v}$ and the vertices of
$V(\vec{v})$.

The thickness $\Theta$ of the lattice covering is given by
\begin{equation}
\Theta(\Lambda)=\frac{\mbox{volume of $d$-dimensional sphere of
radius $R(\Lambda)$}}{\mbox{volume of the elementary cell of
$\Lambda$}}
\end{equation}
The covering problem asks to find a lattice with the lowest thickness.
%
%
The thinnest {\it lattice} coverings are known in dimensions up to
5. They are given by the so called Voronoi's principal lattices of
the first type and are denoted by $A^*_d$. $A^*_2$ is equivalent to
the hexagonal lattice and is proved to be thinnest covering of the
plane, $A^*_3$ is equivalent to the {\it body-centered-cubic} (bcc)
lattice. For the results of the best known coverings in higher
dimensions we refer readers to \cite{Con}.

\subsubsection{Covering problem with constraints}
In our search scheme the calculation of the ${\mathcal F}$-statistic
involves two Fourier transforms that can be computed
efficiently using the fast Fourier transform algorithm. For this
reason we want the nodes of the grid to coincide with Fourier
frequencies: $\Delta\omega, 2\Delta\omega, 3\Delta\omega,\ldots$ for
some fixed frequency resolution $\Delta\omega$. This imposes a
condition that one of the lattice basis vectors has a fixed length
$$
|\vec{l}|=\sqrt{\tilde{\Gamma}\left[(\Delta\omega, 0, \ldots, 0),
(\Delta\omega, 0, \ldots, 0)\right]}
$$
and forbids an immediate use of the general results of the theory of
lattice coverings. Instead one can formulate the {\it covering
problem with constraint}: to find the thinnest lattice covering of
the $d$-dimensional space with spheres of radius $r$ and one of the
basis vectors of the lattice having fixed length $|\vec{l}|$. As far
as we know the general solution to the problem is not known.
We present a construction of a nearly optimal lattice that satisfies the
constraint with a good accuracy.

Let a vector $\vec{v}_0$ define the frequency resolution. We search
for a lattice $\Lambda(w')$ of covering radius $R(\Lambda(w'))=r$
with lattice basis $w'$ satisfying the constraints that can be
expressed as $w'=\left\{ \vec{v}_0,\vec{w}'_1,\ldots,
\vec{w}'_{d-1}\right\}$. We find the thinnest constrained lattice
starting with an optimal unconstrained lattice in $d$-dimensions.
The idea is to shrink the optimal lattice as little as possible such
that one of the basis vectors of the resulting lattice coincides
with the constraint vector $\vec{v}_0$.
We notice that the orientation of the constraint vector $\vec{v}_0$
has no effect on the optimal constrained lattice, it is only the
length of $\vec{v}_0$ and assumed value of covering radius that
matters (and more precisely: only their ratio because the overall
scale can be taken arbitrarily).

For a given lattice $\Lambda$ there always exists the lattice vector
$\vec{l}$ such that $||\vec{v}_0| - |\vec{l}||$ takes minimum value
that we denote by $\vec{l}(\Lambda)$.
We define Algorithm 1:

\hspace{0.4cm}

\begin{footnotesize}
\begin{tabular}{l}
\hline
Listing 1. "Minimal" deformation of a lattice. One of the
nodes\\ of the final lattice coincides with the resolution vector\\
\hline\hline
{\bf Input}: Lattice $\Lambda$; vector $\vec{v}_0$.\\
{\bf Output}: Lattice $\Lambda'$; $\vec{l}(\Lambda')=\vec{v}_0$\\
{\bf A1.} Find $\vec{l}(\Lambda)$.\\
{\bf A2.} Contract $\Lambda$ along $\vec{l}(\Lambda)$ to obtain $\Lambda_c$ with
$|\vec{l}(\Lambda_c)|=|\vec{v}_0|$.\\
{\bf A3.} Rotate $\Lambda_c$ to obtain a lattice $\Lambda_{rc}$ with
$\vec{l}(\Lambda_{rc})=\vec{v}_0$.\\
{\bf A4.} Return $\Lambda'=\Lambda_{rc}$.\\
\hline
\end{tabular}
\end{footnotesize}

\hspace{0.4cm}

For optimal initial lattices Algorithm 1 defines the
following function $f_{\vec{v}_0}:{\bf R}\rightarrow {\bf R}$: for
$x=R(\Lambda)$, $f_{\vec{v}_0}(x)=R(\Lambda')$. For a given
$r\in{\bf R}$ we denote by $r_i$ the value of $x$ for which the
function $|f_{\vec{v}_0}(x)-r|$ reaches its minimum. The optimal
constrained lattice is obtained by application of Algorithm 1
to an optimal (unconstrained) lattice $\Lambda$ with
covering radius $R(\Lambda)=r_i$.

\begin{figure}[htp]
\includegraphics[height=22pc,width=22pc]{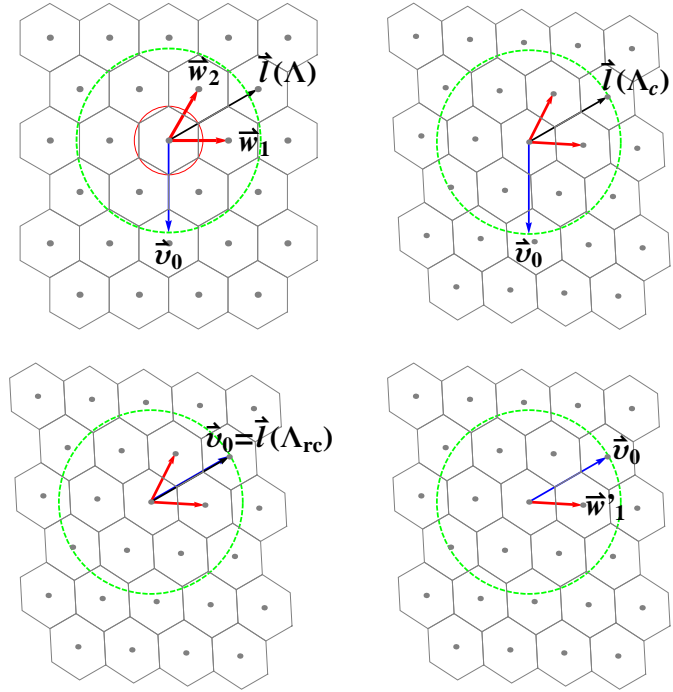}
\caption{Illustration of Algorithm 1 in two dimensions.
Initial optimal lattice (top left) with the basis
$\{\vec{w}_1,\vec{w}_2\}$, resolution vector $\vec{v}_0$ defining
constraint surface and vector $\vec{l}$ is contracted along the
vector $\vec{l}$ (top right) and rotated (bottom left) such that for
the new lattice $\vec{l}=\vec{v}_0$. Final suboptimal lattice has
basis $\{\vec{v}_0,\vec{w}_1'\}$ (bottom right).}
\label{fig:alg1234}
\end{figure}

In dimensions $d=2,3,4,5$ as the initial lattices one takes $A^*_d$
lattices but the procedure can be generalized to any number of
dimensions by taking as the input the best known lattice covering in
a given dimension \cite{Con}. Fig. $\ref{fig:alg1234}$ illustrates
the procedure for two--dimensional $A^*_2$ hexagonal lattice. It is
seen there that contraction of the lattice can change the initial
Voronoi cell and covering radius.
However when the vector $\vec{l}(\Lambda)$ initially lies on the
"constraint surface" depicted by the dashed circle in the Fig.
$\ref{fig:alg1234}$ the procedure acts trivially (no contraction only
rotation) leaving the final lattice optimal. One often encounters
trivial procedures when the vector $\vec{l}(\Lambda)$ is large as
compared to the Voronoi cell and moreover in these cases contractions
are small. On the other hand the last trivial procedure occurs when
the resolution vector and the shortest lattice vector have equal
lengths.

\begin{figure}[htp]
\begin{center}
\subfigure{
\includegraphics[width=35pc]{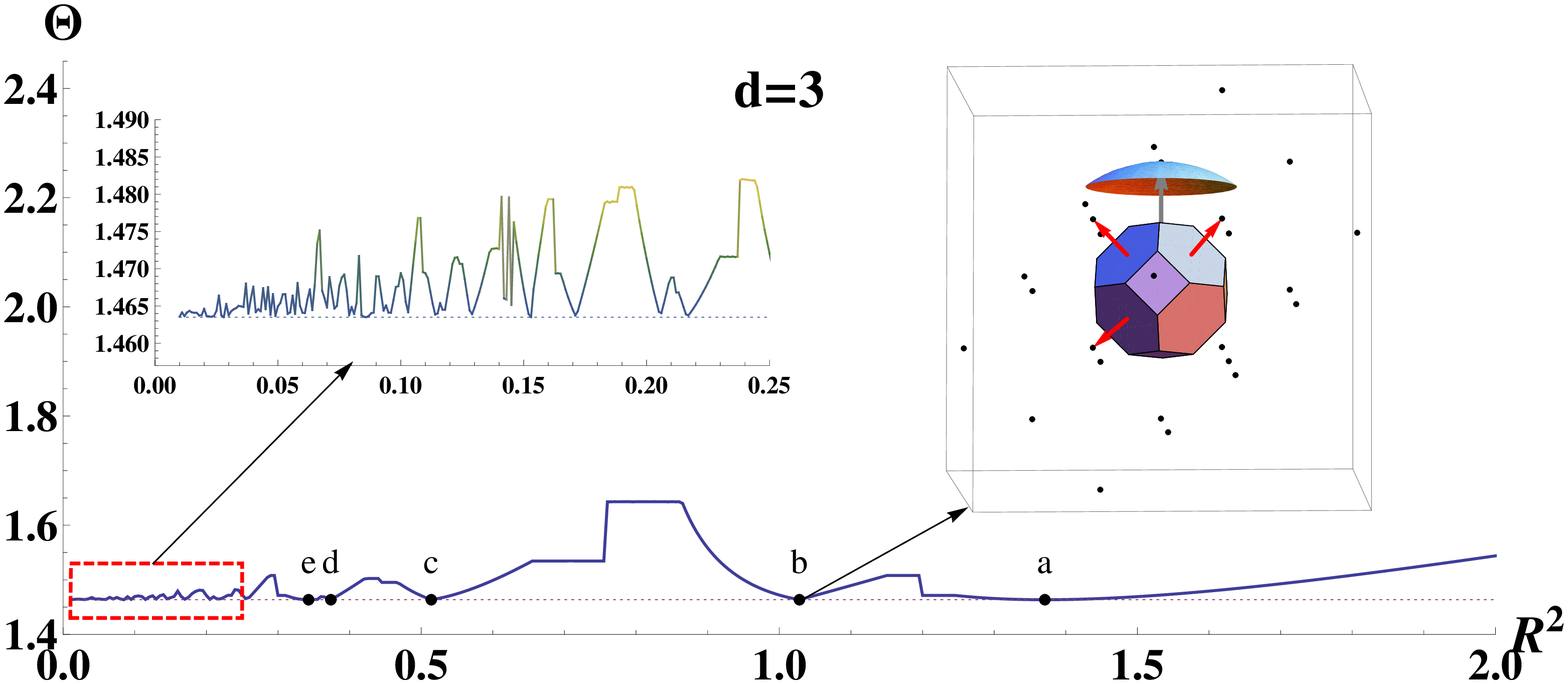}}
\subfigure{
\includegraphics[width=35pc]{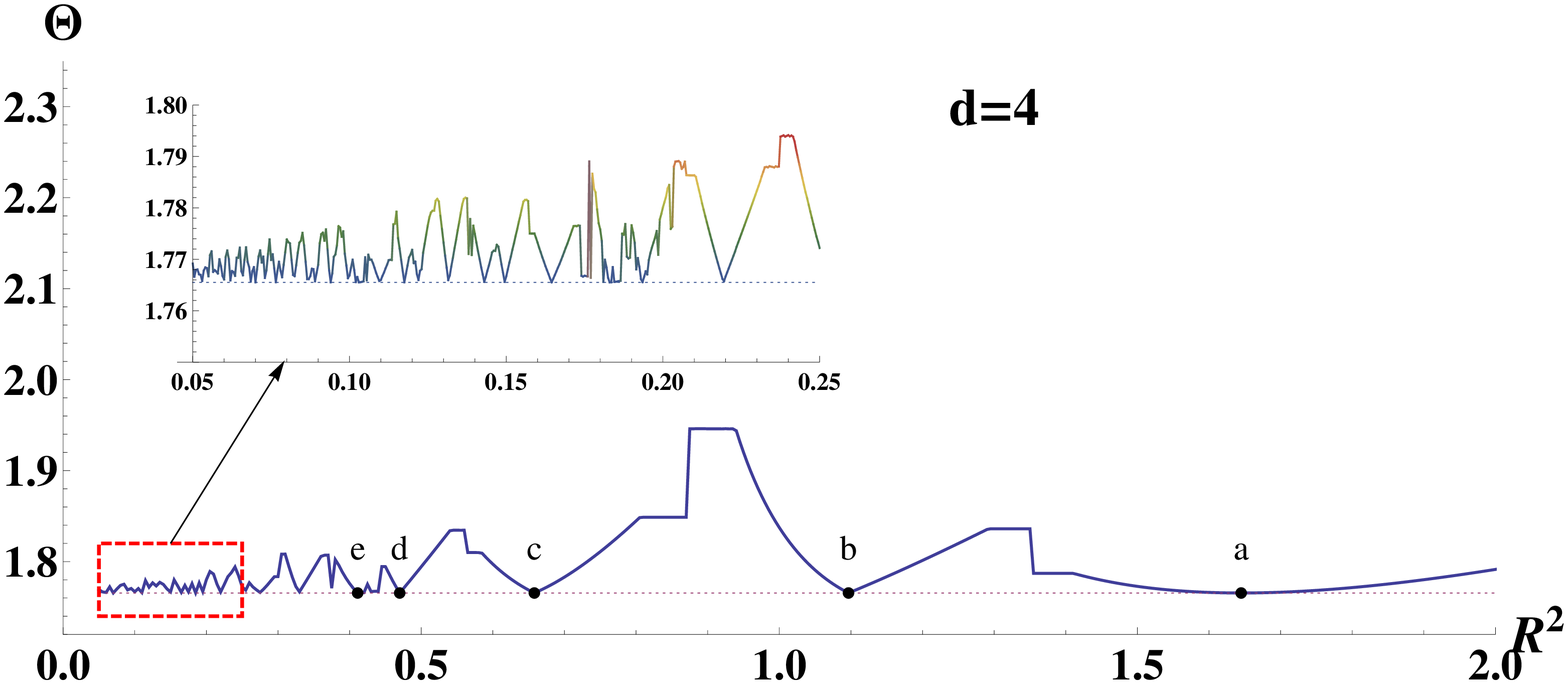}}
\end{center}
\caption{Thickness of the lattices in 3 and 4 dimensions compared to
the optimal thickness of $A^*_3\approx 1.4635$ and $A^*_4\approx
1.7655$. The upper diagram shows also the optimal lattice with
covering radius $R(A^*_3)^2=5/48\pi^2$. The Voronoi cell, basis
vectors, resolution vector and part of the constraint surface are
depicted; for this specific value of $R$ the head of the resolution
vector lies on the constraint surface.} \label{fig:thr2d34}
\end{figure}

One can determine the values of covering radii for which the final
lattices are optimal. As an example we find the values of four
largest covering radii having this property. To do this we consider
the sequence of the shortest vectors of the lattice. In three and
four dimensions the first nonzero shortest vectors of $A^*_3$ and
$A^*_4$ lattices have lengths $2\sqrt{3/5}R(A^*_3)$,
$4/\sqrt{5}R(A^*_3)$, $4\sqrt{2/5}R(A^*_3)$, $2\sqrt{1/11}R(A^*_3)$,
$\ldots$ and $\sqrt{2}R(A^*_4)$, $\sqrt{3}R(A^*_4)$,
$\sqrt{5}R(A^*_4)$, $\sqrt{7}R(A^*_4)$, $\ldots$ respectively. The
length of the resolution vector which in dimensionless units has the
from $\vec{v}_0=(2 \pi,0,\ldots)$ is equal to
$\sqrt{\tilde{\Gamma}(\vec{v}_0,\vec{v}_0)}=\pi/\sqrt{3}$ in both
dimensions. This gives the following squares of the largest covering
radii for optimal constrained lattices: $R(A^*_3)^2$:
$5/36\pi^2\approx 1.3708$, $5/48\pi^2\approx 1.0281$,
$5/96\pi^2\approx 0.514$, $5/132\pi^2\approx 0.3739$, $\ldots$ and
$R(A^*_4)^2$: $\pi^2/6\approx 1.6450$, $\pi^2/9\approx 1.0966$,
$\pi^2/15\approx 0.6580$, $\pi^2/21\approx 0.4700$ [points a, b, c,
d on Fig.(\ref{fig:thr2d34})]. For larger values of $R$ thickness of
final lattices grows monotonically with covering radius.

These features are seen in the Fig. $\ref{fig:thr2d34}$  which
shows the results of application of the construction to the model
(\ref{sm}) in 3 and 4 dimensions for different covering radii. The
resolution vector is $\vec{v}_0=(2 \pi,0,\ldots)$ and the
observation time is 2 years.
As for a fixed volume of the parameter space the number of points
needed to cover the space with balls of a given radius (i.e. for allowed loss
of signal-to-noise ratio)
is proportional to the thickness, the diagram demonstrates that the
excess of points due to constraints is minor for the wide range
of radii which makes the search strategy based on FFT effective.

\subsection{Number of false alarms}
\label{subsec:NF}
In order to estimate the number of false alarms expected in our
search we use a general approach consisting of dividing the
parameter space into elementary cells defined by the autocorrelation
function of the ${\cal F}$-statistic (\cite{JK} and \cite{JKbook}
Chapter 6.1.3). We use the Taylor expansion of the autocorrelation
function around the true values of the parameters $\theta_k$
and moreover we use an approximate response of the detector
given by the linear model (Eq.\, (\ref{smn}) or (\ref{smn3})). In
this case the hypervolume of the elementary cell is given by the
volume $V_c$ of the hyperellipsoid defined as
\begin{equation}
\tilde{\Gamma}_{kl}\Delta\theta_k\Delta\theta_l \leq 1/2.
\end{equation}
Thus $V_c$ is given by
\begin{equation}
\label{eq:vc} V_c =
\frac{(\pi/2)^{m/2}}{\Gamma(m/2+1)\,\sqrt{\det\tilde{\Gamma}}},
\end{equation}
where $m$ is the dimension of the parameter space and $\Gamma$
denotes the Gamma function. The determinants of the reduced Fisher
matrix in 3 and 4 dimensional cases read
\begin{equation}
\det\tilde{\Gamma_3} =  \frac{T_o^2}{48}\frac{\pi^2 n^2  - 6}{\pi^2
n^2}
\end{equation}

\begin{equation}
\det\tilde{\Gamma} = \frac{T_o^6}{34560}\frac{(\pi^2 n^2  - 6)(\pi^4
n^4 - 90)}{\pi^6 n^6},
\end{equation}
where it is assumed that the observation time $T_o$ is an integer multiple $n$ of years. As
for the linear model the reduced Fisher matrix has components
independent of the values of the parameters and the number of cells is
simply given by
\begin{equation}
\label{NC} N_c = \frac{V}{V_c},
\end{equation}
where $V$ is the hypervolume of the parameter space. We assume that
we search a narrow band of bandwidth $\Delta\omega$ with upper
frequency $\omega_\mathrm{max}$. Then the volume of the parameter
space $V_3$ when the frequency drift is not included in the search
is given by
\begin{equation}
\label{eq:V3} V_3 = 2 \Delta\omega\,\pi\, \omega_\mathrm{max}^2  R^2
\end{equation}
whereas the hypervolume $V_4$ when it is included is given by
\begin{equation}
V_4 = V_3\,\Delta\dot{\omega},
\end{equation}
where
$\Delta\dot{\omega}$ is the range of the frequency drift parameter
$\dot{\omega}$. The factor of 2 in Eq.\,(\ref{eq:V3}) is because we
search the space spanned by parameters $\omega$, $\dot{\omega}$,
$A$, and $B$ twice - both for positive and negative values of the
ecliptic latitude $\beta$.

The expected number of the false alarms $N_F$ is given by
\begin{equation}
\label{NF}
N_F = N_c P_F(\Fo),
\end{equation}
where $P_F$ is the probability of false alarm. In the case of linear
model $P_F$ is the $\chi^2$ probability distribution with two
degrees of freedom i.e.
\begin{equation}
P_F(\Fo) = \exp (-\Fo)
\end{equation}

\subsection{Computation of the $\F$-statistic}
\label{subsec:Fcomp} The ${\mathcal F}$-statistic is computed
approximately taking advantage of the speed of the FFT algorithm as
described in Section \ref{subsec:FFT}. The ${\mathcal F}$
is calculated on the grid constructed in Section \ref{subsec:grid}.
For each parameter pair $(A,B)$ the ${\mathcal F}$-statistic is
computed for both positive and negative value of the ecliptic
latitude $\beta_{\rm mid}$ (see Eq.\,(\ref{eq:bemid})). Approximate
calculation of the $\F$-statistic on the grid described
above is called the {\em coarse search}. Using the coarse search we
identify signals for which the ${\mathcal F}$-statistic crosses a
certain threshold. The coarse search is then followed by the second
step which we call the {\em fine search}. Fine search consists of
a search for the maximum of the ${\mathcal F}$-statistic around the
parameters identified by the coarse search. To find the maximum we
use the Nelder-Mead maximization algorithm \cite{LRWW98}. In the
fine step we use accurate expressions  for the ${\mathcal F}$-statistic
[Eq.\ (\ref{eq:mld3}) or (\ref{eq:mld3N})], without
approximations described in Section \ref{subsec:FFT}. As initial
values for the Nelder-Mead algorithm we use the parameters obtained
in  the coarse search. The values of the parameters corresponding to
the maximum of ${\mathcal F}$-statistic are our final estimates of
the parameters of the signal.

\section{Search strategy}
\label{sSearch}
In our entry for challenge $3.1$ we have used the
following procedure to extract GW signals from white dwarf/white dwarf
binaries in the mock LISA data. We search the band from frequency $f = 0.1$ mHz to
frequency 12 mHz where $f=\omega/2\pi$. We do not
go to higher frequencies because with our search strategy it would
involve much more computing time than we could afford. Also above
12 mHz the expected number of white dwarf binaries in our Galaxy is very small. We first
divide the data into bands of 0.1 mHz each. To obtain narrowband
data in the frequency band $[f_1 \,\,\, f_2]$ we first pass the data
through 3rd order Butterworth filter with passband of $[f_1-\epsilon
\,\,\, f_2+\epsilon]$  where we choose the edge parameter $\epsilon$
equal to 0.005 mHz. Then we shift the data to DC by frequency
$f_1-\epsilon$ and we pass it again through the 3rd order
Butterworth filter with passband of $[ f_2 - f_1 + 2\epsilon]$.
After each Butterworth filter we downsample the data. This reduces
the number of data points by a factor of around 300. Each
Butterworth filter is applied twice: forward and backward in time.
In this way there is no phase shift of the narrowbanded data with
respect to the original one. In the search of each narrow band for signals
we neglect the edges of the band of $\epsilon = 0.005$ mHz and
therefore we only search the band $[f_1 \,\,\, f_2]$.
We have included in our search the frequency drift parameter
$\dot{\omega}$. Analysis of the accuracy of estimation of $\dot{\omega}$
have shown (see Section \ref{eqc:res} below) that it is useful to include
the $\dot{\omega}$ parameter in the search for frequencies above $3$mHz.
We have selected the range of the $\dot{\omega}$
parameter using a fit from the values of the $\dot{\omega}$
parameter in the key of the challenge 3.1 training data set. In each band we
search for the signals calculating the ${\cal F}$-statistic over the
constrained grid constructed in Section \ref{subsec:grid}. This
enables application of the FFT algorithm in $\F$-statistic
computation. We select the strongest signal and to this signal we
apply the fine search described in Section \ref{subsec:Fcomp} to
estimate its parameters. We use the Nelder-Mead algorithm  with the
initial values provided by the parameters of the template with the
largest value of the ${\cal F}$-statistic over the grid. We
reconstruct the signal in the time domain and remove it from the
data. We then search for the next strongest signal and so on until
the signal-to-noise ratio (SNR) of the detected signal estimated as
\begin{equation}
\label{eq:snr}
SNR=\sqrt{2({\mathcal F}-2)}
\end{equation}
falls below a certain threshold.
We have chosen the threshold for the ${\cal F}$-statistic equal to 18. This corresponds to SNR threshold of
around 5.7 (see Eq.\,(\ref{eq:snr})). The number of signals increases as the signal-to-noise ratio
decreases. At a sufficiently low signal-to-noise ratio the signals are so close to each other
in the parameter space that they interfere with each other. The $\F$-statistic (Eq.\,\ref{eq:mld3N})
that we use in our search was derived under the assumption that there was only one signal present in the data so it can be used to detect multiple signals
when they are sufficiently separated in the parameter space so that the $\F$-statistic for the multiple signals is a sum of the
${\cal F}$-statistics for individual signals.
Thus when there are many signals present we neglect the interference between the signals. Our choice
of 18 of the threshold for the ${\cal F}$-statistic was a convenient choice that, as we shall see
in the following section, led to a detection and accurate estimation of over $10^4$ signals.

\begin{figure}[h]
\includegraphics[width=25pc]{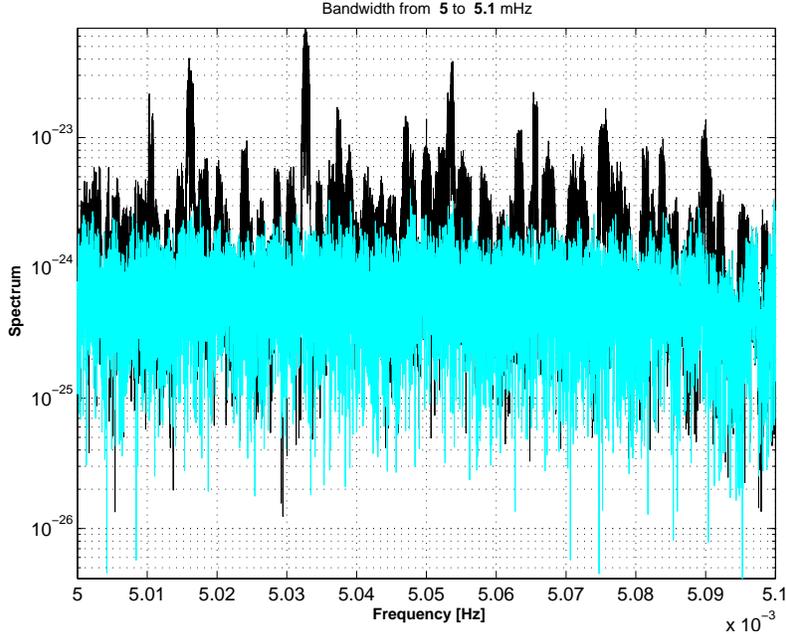}
\caption{\label{res50} Estimation of the signals from white dwarf
binaries in the challenge 3.1 data set for the band from 5 mHz to 5.1 mHz.
Black color denotes the original data and the light blue color is the data after signals removed.}
\label{re1}
\end{figure}

\begin{figure}
\includegraphics[width=25pc]{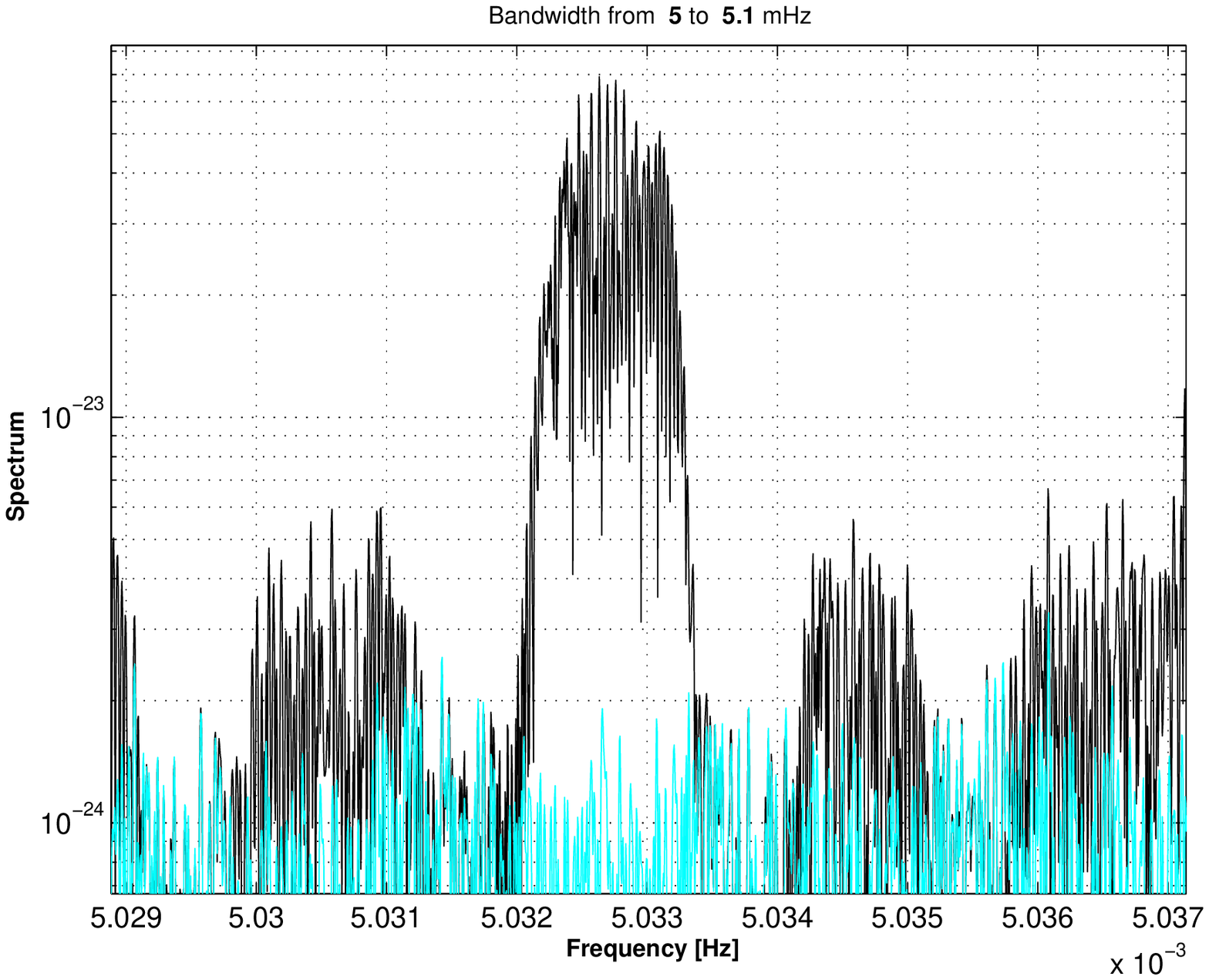}
\caption{\label{res50strongest} The zoom of the Fig. \ref{res50}
around the strongest signal identified.}
\label{re2}
\end{figure}

In the Figs. $\ref{re1}$ and $\ref{re2}$ we have presented
application of the above strategy to the challenge 3.1 data
set in the bandwidth from 5 mHz to 5.1 mHz.

In this bandwidth we have identified 132 signals altogether out of 168 present. The
accuracy of estimation of most of the signals is one sigma where
sigma is calculated from the Fisher matrix. For several signals the
error was very large indicating that either noise mimicked the signal
or the residuals of removed signals remained significant.
This could happen as a result of interference of the signals.

\section{MLDC results}
\label{eqc:res}
Here we present results of our entry for challenge $3.1$
\cite{web:mldc}. The challenge $3.1$ consisted of a two-year data
set with $15$ s sampling time with signals from around $6\times
10^7$ binaries. Our aim was to detect as many as possible from the
40628 brightest binaries present in the data.
We have performed a self-evaluation of our search by the
following procedure. In our procedure we have used the correlation
$C(s_1,s_2)$ between the two signals $s_1$ and $s_2$ defined as
%
%
\begin{equation}
\label{eq:corrf}
C(s_1,s_2):=\frac{\av{s_1,s_2}_{\cal
S}}{\sqrt{\av{s_1,s_1}_{\cal S}}\sqrt{\av{s_2,s_2}_{\cal S}}}.
\end{equation}
The first step of our self-evaluation consisted
of selection of the true detected signals from the set of all
submitted data. For a submitted signal $s$ with parameters
$\theta_s$ we considered set ${\bf B}_s$ of all key signals within
the frequency bins about $f_s$, i.e. signals satisfying
$$
|f_s-f_k|<1/T_{obs}\approx 1.6 \times 10^{-8}\mbox{Hz},
$$
where parameters $\theta_k$ of the key signals were taken from the
set of $40628$ bright Galactic binaries \cite{web:mldc}. Next, each
signal $s$ was paired with the key signal $k_s$ from ${\bf B}_s$
that maximizes the correlation:
\begin{equation}
s \mapsto k_s: \;\;\;\;\;\;k_s=\arg\max_{k\in{\bf
B}_s}C\left(s,k\right)\,;
\end{equation}
we interpreted $\theta_{s}$ as the parameters estimates of the key
signal $k_s$. In the case of multiple detection of a key signal
$k_0$, that is when $k_{s_1}=k_{s_2}=\ldots=k_0$ for different
$s_1$, $s_2$, $\ldots$, as the main signal $s_{k_0}$ we singled out
the one that maximizes the correlation with $k_0$,
\begin{equation}
s_{k_0} = \arg{\max_{s:\, k_s=k_0}} C\left(s,k_0\right),
\end{equation}
and we rejected the remaining secondary signals.

\subsection{Original results}
\begin{figure}
\includegraphics[width=40pc]{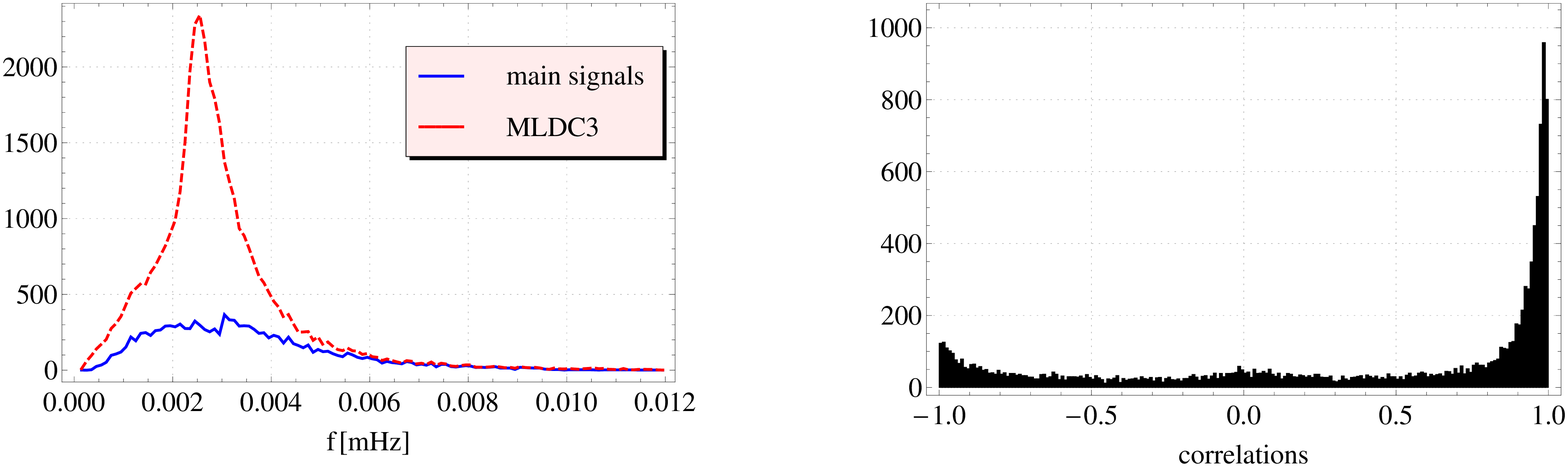}
\caption{\label{detcorr:old} Detection and
correlations for the blind challenge $3.1$ data set in the original run.
The left panel shows number of the signals detected by our search and selected by
the procedure described in the text as a function of the frequency.
The right panel displays the histogram of the correlation functions
(Eq.\,(\ref{eq:corrf})) between our estimated signals and the ones
form the key.}
\end{figure}

Our original entry for
challenge $3.1$ contained a bug in our data reading procedure.
Our analysis for frequencies above $3$mHz was performed on the challenge data set shifted
in time by $1440$s. In the original search we have
detected 14838 signals altogether and we have estimated their
parameters.
Results of detection are displayed in Fig. \ref{detcorr:old}
which shows detected (main) signals and the correlations with signals from the key.
We see that the histogram of the correlations shows an excess
of anticorrelations due to the bug in our
data reading procedure.
\begin{figure}
\includegraphics[width=25pc]{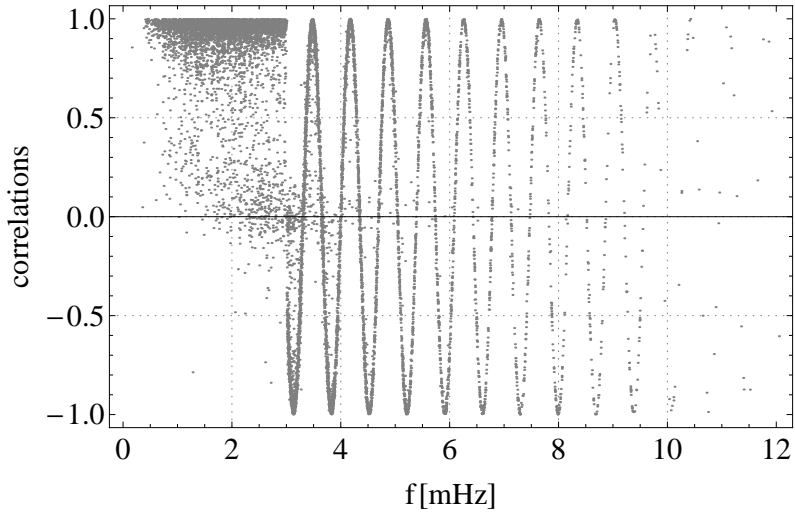}
\caption{\label{corr:old} Correlations vs. frequency.
The periodic variation of the correlation
function was due a time shift in reading of the challenge data set.}
\end{figure}
Effects of this systematic error can be seen in the Fig.
\ref{corr:old} displaying correlations as the function of the
frequency. Peculiar oscillations of the correlation with frequency
are easily explained by noticing that the phase difference between
nearly monochromatic signal and the same signal shifted in time by
$t_0$ is equal to $2\pi f t_0$ and gives periodically changing
overlap between the two functions which oscillates with respect to
$f$ with "frequency" $t_0$ or by explicitly computing correlation of
two monochromatic waves $\cos{[2\pi f(t-t_0)]}$ and $\cos{(2\pi
f)}$. This explains the excess of signals with anticorrelations in
Fig. \ref{detcorr:old}.

\subsection{Corrected results}
We have corrected the bug in the reading procedure and have made a
second run of the challenge $3.1$ data set.
Here we present results of this second run.
In our search we have detected 17051 signals altogether and we have estimated their
parameters.

In the band from 0.1mHz to
3mHz we have made two runs - one with and the other without the
$\dot{\omega}$ parameter included. We have used these two runs to investigate at what
frequency it is useful to start estimating the frequency drift
of the GW signal. In the search including the  $\dot{\omega}$
parameter, we have found that we can start estimating it only above
frequency of 0.5 mHz because below this frequency the cell of our
4-dimensional grid is bigger than our parameter space. Including the
frequency derivative from 0.5mHz we have identified  16785 signals
and including $\dot{\omega}$ starting only from 3mHz we have found 17051 signals.
We have compared the absolute values of the errors in $\dot{\omega}$
when we include it in the search and when we do not include it. When
we do not include it we take as the error the absolute value of
the $\dot{\omega}$ parameter from the challenge key signal. We calculate
the mean values of the these absolute errors in each band. We find
that slightly below the frequency of 3mHz the mean error for the
case when we include $\dot{\omega}$ parameter becomes less than when
we do not include it. Thus we find that our initial choice of
threshold frequency equal to 3 mHz to include the $\dot{\omega}$ is a
reasonable one. Therefore, as our final result, we consider the signal
found using the search that turns on the frequency derivative at
3mHz frequency. We have also estimated the expected number of false
alarms using the formula (\ref{NF}) in Section \ref{subsec:NF}. We
find that for low frequencies where we detect most of our signals
the number of false alarms is negligible. The number of false alarms
increases quadratically with frequency and linearly with the range
of $\dot{\omega}$ parameter. The expected number of false alarms
exceeds one only at the frequency $f = 8$ mHz.

\begin{figure}
\includegraphics[width=40pc]{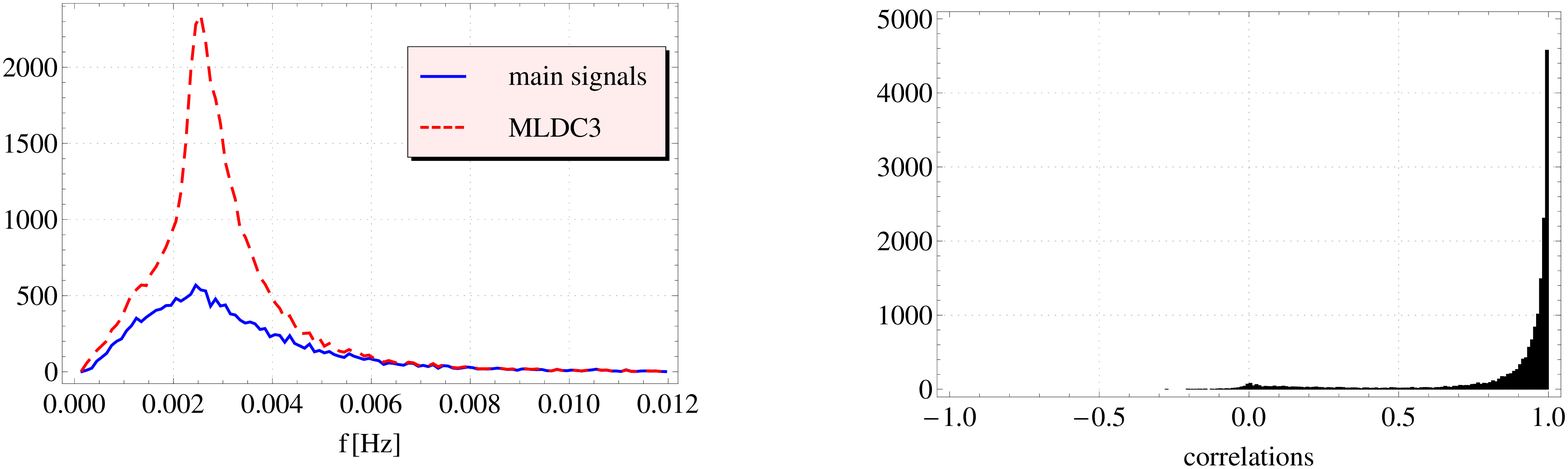}
\caption{\label{detcorr:new}  Detection and
correlations for the blind challenge $3.1$ data set in the second run.}
\end{figure}

\begin{figure}
\includegraphics[width=30pc]{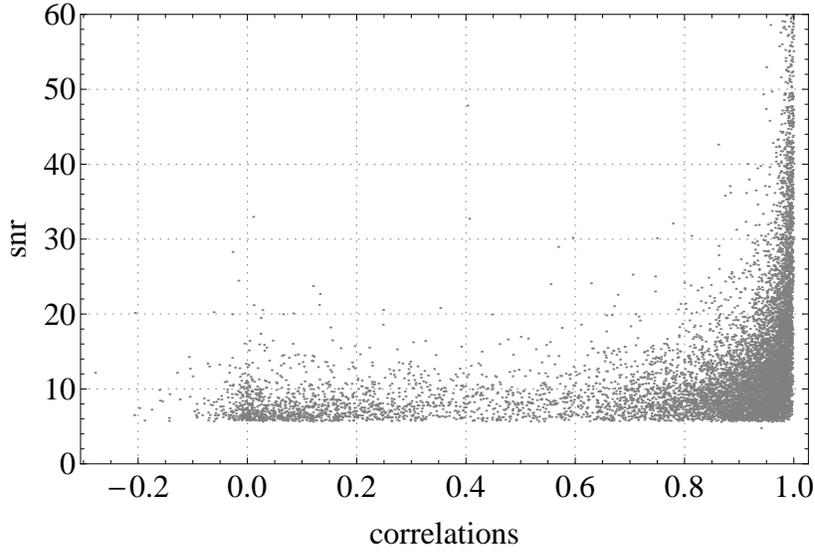}
\caption{\label{fig:corrSNR} Number of correlations of estimated
signals with key signals as a function of the signal-to-noise ratio
for signals with frequency below 3 mHz.}
\end{figure}

In Fig. \ref{detcorr:new} we present the number of signals
detected against all the challenge $3.1$ signals and the
correlations of the estimated signals with the key signals
in the second run.
We see that there is still an excess of correlations with
correlation parameter around zero. We have investigated the number
of correlations as a function of signal-to-noise ratio (see Fig.
\ref{fig:corrSNR}).
We find that the excess of low correlations originates from frequencies
below 3 mHz. Moreover for SNR $> 10$ the number of small
correlations considerably decreases. The number of selected signals
after we discarded signals below SNR $= 10$ and for frequency less than 3
mHz is 12805. We quote this number as the number of signals which we
detect and accurately estimate the parameters.
We need to refine our data analysis methods in order to extract reliable
estimates of the parameters
for the signals of SNR less
than 10 and frequency less than 3 mHz.

\begin{figure}
\includegraphics[width=40pc]{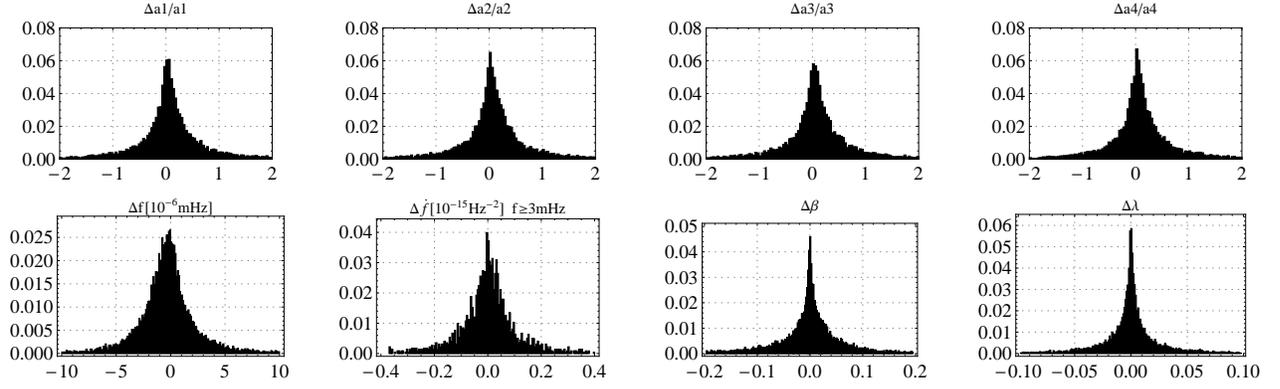}
\caption{\label{errorsnew} Errors of the parameters of the signals detected and verified in our
search of the challenge $3.1$ data set.}
\end{figure}

\begin{figure}
\includegraphics[width=40pc]{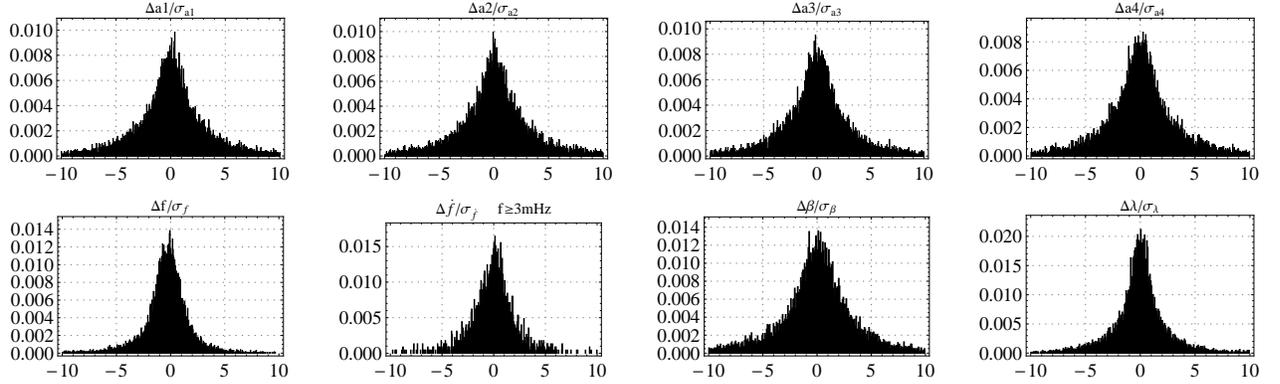}
\caption{\label{CRerrors} Errors of the estimation of parameters as in Fig.
\ref{errorsnew} but normalized by the variances obtained from the inverse of the
Fisher information matrix.}
\end{figure}

\begin{figure}
\includegraphics[width=35pc]{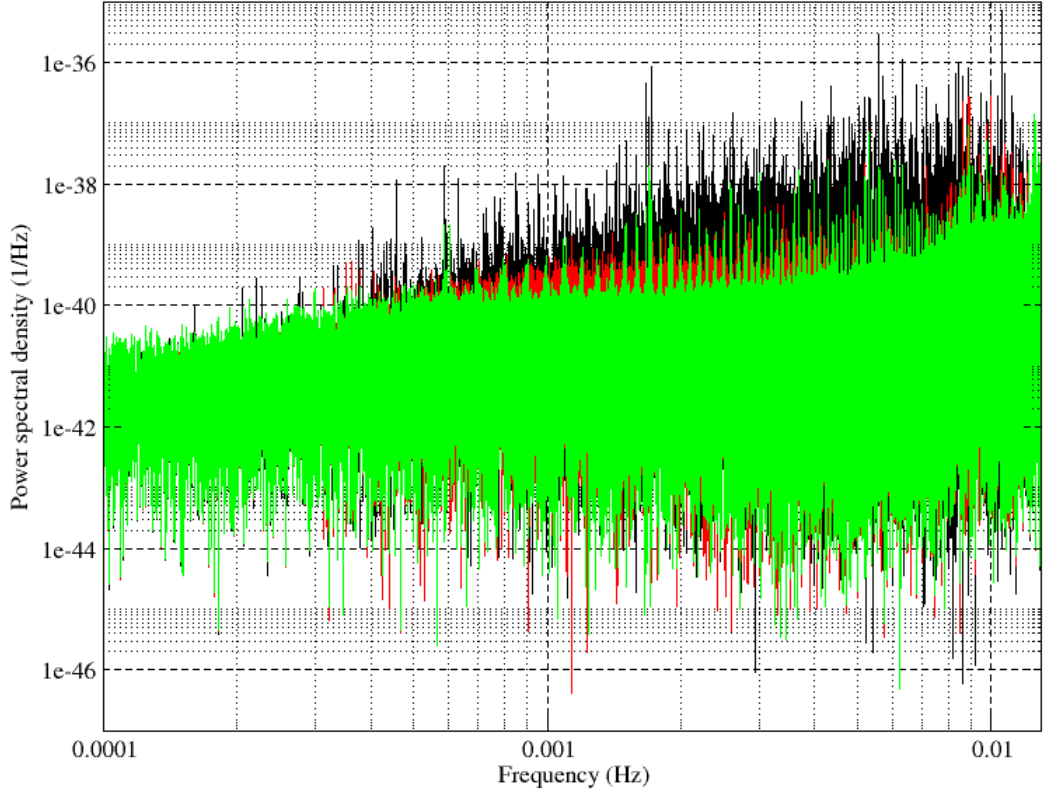}
\caption{\label{ReductionFFT} Amplitude square of the Fourier transformed data.
Upper signal is original challenge data (A-channel, includes full
Galaxy and instrumental noise), middle is the data with selected
removed signals, and lower is the data with all identified signals
removed.}
\end{figure}

\begin{figure}
\includegraphics[width=35pc]{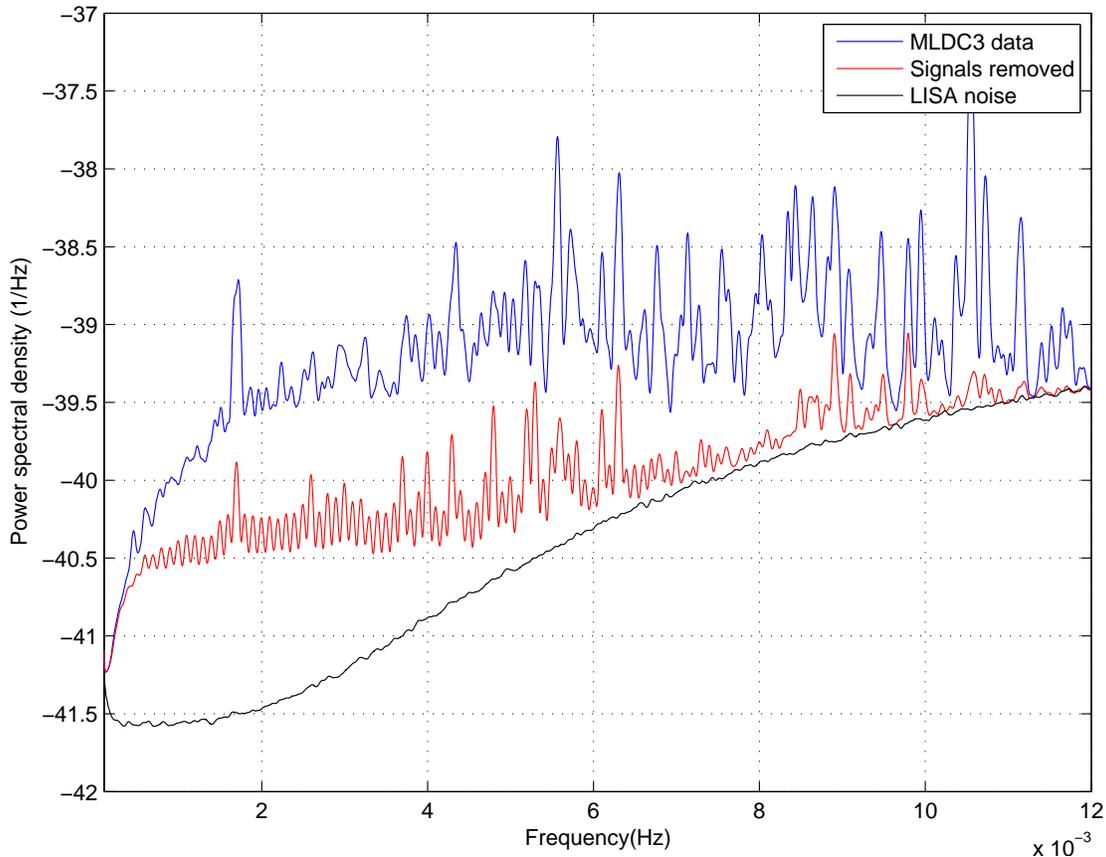}
\caption{\label{ReductionSpec} Power spectral density.
Upper signal is the original challenge data,
middle is reduced data (after removing all found signals),
and lower is the instrumental noise.
}
\end{figure}

The excess of zero correlation signals arises
either because for some low signal-to-noise ratio the
parameter estimation was not accurate or/and  because at low
signal-to-noise ratio there are much more signals that interfere
causing biases in the parameter estimators. Fig. \ref{errorsnew}
demonstrates the results of the parameter estimations for
our search of challenge 3.1 data set. Errors are defined
as differences between the key and the recovered signal parameters,
$\Delta\theta:=\theta_{key}-\theta_{rec}$. Histograms in Fig.
\ref{CRerrors} show the parameter estimation errors divided by the
standard deviations $\sigma_{\theta_{key}}$ obtained from the Fisher
matrix for the key signals.
In Fig. \ref{ReductionFFT} we have presented the power spectrum
of the challenge 3.1 data against the power spectrum of the data
after the detected signals were removed. We plot two power
spectra: one when all the signals identified are removed and the
other one when only the ones selected by our procedure are removed
form the challenge data set. We notice two effects. One is that
periodically our identification procedure gets worse. This because
we were not estimating parameters of the signal very well at the
edges of the narrow bands. We have found that this is because our
narrowband filtering procedure was not perfect.
The other effect is that when we remove all identified signals
(lower signal on the Fig. \ref{ReductionFFT}) we are doing
a much better job then when we remove only signals identified
as true signals by our procedure (middle signal
on the Fig. \ref{ReductionFFT}). Thus we fit quite a
large number of signals well but with wrong parameters.
In Fig. \ref{ReductionSpec} we have compared smoothed spectrum of the challenge 3.1 data set
with that of data with identified signals removed and we have compared them with
spectrum of the LISA detector instrumental noise.
From this figure we conclude that above frequency of $6$ mHz we resolve all the
white-dwarf binary systems well.

\section{Conclusion}
\label{sec:Con}
From the analysis of our challenge $3.1$ results we see that in order to increase substantially
the number of signals with good parameter estimation we need to assume in deriving our
filters that there are more than one signal present in the data. However even with the
current procedure that estimates signals one by one we can still make the following improvements.

\begin{enumerate}

\item
Improve the splitting of the time series into narrow bands.

\item
Lower the threshold for detection.

\item
For high frequencies bands where signals are well separated identify all the signals
at one scan of the parameter space instead extracting only one and scanning the whole
parameter space again to go to the next.

\end{enumerate}

\appendix
\section{Linear model}
\label{app:lm}
In this appendix we consider the {\em linear model}, an approximation
of the full gravitational wave signal introduced in order to construct
a metric and a grid on the intrinsic parameter space. It has the form:
\begin{equation}
\label{sm}
s(t)= A_0\cos{(\omega
t+\frac12\dot{\omega}t^2+A\cos{\Omega t}+B\sin{\Omega t}+\phi_0)},
\end{equation}
with constant amplitude $A_0$, constant phase $\phi_0$
and two parameters $A$, $B$ (related to $\omega$, $\beta$ and $\lambda$).

Following the general case (\ref{X}) we rewrite the signal
(\ref{sm}) in the form
\begin{equation}
\label{sma} s(t)= A_1 X_1(t) + A_2 X_2(t),
\end{equation}
where
$$
X_1=\cos{(\omega t+\frac12\dot{\omega}t^2+A\cos{\Omega
t}+B\sin{\Omega t})},
$$
$$
X_2=\sin{(\omega t+\frac12\dot{\omega}t^2+A\cos{\Omega
t}+B\sin{\Omega t})}
$$
and where we have introduced the extrinsic amplitudes $A_1=h_0\cos{\phi_0}$
and $A_2=-h_0\sin{\phi_0}$.

The Fisher matrix,
\begin{equation}
\Gamma_{ij}=\frac{2 T_o}{S(\omega_o)}\av{\partial_i s\;\partial_j
s},
\end{equation}
for the linear model (\ref{sma}) with respect to the parameters
$(A_1, A_2, \omega, \dot\omega, A, B)$ takes the form
\begin{equation}
\label{fm6} \Gamma =
\left(%
\begin{array}{cc}
   {\bf G}_1 & {\bf G}_2 \\
   {\bf G}_2{}^T & {\bf G}_3
\end{array}%
\right),
\end{equation}
where
\begin{eqnarray}
\label{fm61}
{\bf G}_1 & = & \rho^2\left(%
\begin{array}{cc}
   \frac{1}{A_1{}^2+A_2{}^2} & 0 \\
   0 & \frac{1}{A_1{}^2+A_2{}^2}
\end{array}%
\right) \\
\label{fm62}
{\bf G}_2 & = & \rho^2\left(%
\begin{array}{cccc}
\frac{T_o{}A_2}{2(A_1{}^2+A_2{}^2)} & \frac{T_o{}^2 A_2}{6(A_1{}^2+A_2{}^2)} & 0 & 0 \\
-\frac{T_o{}A_1}{2(A_1{}^2+A_2{}^2)} & -\frac{T_o{}^2 A_1}{6(A_1{}^2+A_2{}^2)} & 0 & 0 \\
\end{array}%
\right) \\
\label{fm63}
{\bf G}_3 & = & \rho^2\left(%
\begin{array}{cccc}
\frac{T_o{}^2}{3} & \frac{T_o{}^3}{8} & 0 & -\frac{1}{\Omega} \\
\frac{T_o{}^3}{8} & \frac{T_o{}^4}{20} & \frac{1}{\Omega^2} & -\frac{T_o}{2\Omega} \\
0 & \frac{1}{\Omega^2} & \frac12 & 0 \\
-\frac{1}{\Omega} & -\frac{T_o}{2\Omega} & 0 & \frac12
\end{array}%
\right),
\end{eqnarray}
and where the optimal signal-to-noise ratio,
\begin{equation}
\label{rhosm} \rho^2 = \frac{2 T_o}{S(\omega_o)}\av{s^2},
\end{equation}
is given by $\rho^2 = \frac{T_o (A_1{}^2+A_2{}^2)}{S(\omega_o)}$ and
the observation time $T_o$ is assumed to be an integer $n$ of years.
In the derivation of Eqs. (\ref{fm61}) - (\ref{rhosm}) we have used
the following approximations
\begin{equation}
\av{X_1{}^2}\simeq\av{X_2{}^2}\simeq\frac12,
\;\;\;\;\;\;\av{X_1\,X_2}\simeq 0
\end{equation}
corresponding to approximations (\ref{eq:app0})-(\ref{eq:P}) for the
case of the full signal.

The reduced Fisher matrix obtained from the full Fisher matrix $\Gamma$ by projecting
$\Gamma$ on the intrinsic parameters space and normalizing it is explicitly given by (\cite{JK})
\begin{equation}
\tilde{\Gamma} = \frac{1}{\rho^2}\left( {\bf G}_3 -{\bf G}_2{}^T \,
{\bf G}_1{}^{-1}\,{\bf G}_2 \right)
\end{equation}

The coefficients of the reduced Fisher matrix $\tilde{\Gamma}$ of
the linear model (\ref{sma}) in the dimensionless units with $T_o=1$
are given by
\begin{equation}\label{appFM4red}
\tilde{\Gamma} =
\left(%
\begin{array}{cccc}
  \frac{1}{12}              & \frac{1}{24}          & 0                     & -\frac{1}{2 \pi n}   \\
  \frac{1}{24}              & \frac{1}{45}          & \frac{1}{4\pi^2 n^2}  & -\frac{1}{4 \pi n} \\
  0                 &  \frac{1}{4\pi^2 n^2}         & \frac12 & 0   \\
 -\frac{1}{2\pi n}  & -\frac{1}{4 \pi n}            & 0   & \frac12 \\
\end{array}%
\right)  .
\end{equation}

For a network of detectors ${\bf s}=(s_A,s_E,s_T)$ with uncorrelated
noises the Fisher matrix and the optimal signal-to-noise ratio can
be written in terms of the noise-weighted averaging operator and
vectorial response [see Sect.(\ref{sec:filter})]:
\begin{equation}
\label{fm6vec1} \Gamma_{opt\, ij} = \frac{2 T_o}{{\cal
S}(\omega_o)}\av{\partial_i{\bf s}\,\partial_j{\bf s}}_{\cal S},
\end{equation}
\begin{equation}
\label{rhosmvec1} \rho_{opt}{}^2 = \frac{2 T_o}{{\cal
S}(\omega_o)}\av{{\bf s}^2}_{\cal S}.
\end{equation}
In the case of network of LISA detectors the optimal responses $A$,
$E$, and $T$ can be approximated by the linear model of the form
\begin{equation}
\label{smn}
s_I(t)= c_I h_0\cos{(\omega
t+\frac12\dot{\omega}t^2+A\cos{\Omega t}+B\sin{\Omega
t}+\phi_0+d_I)},\;\;\;\;\;\;\;\;I=A,E,T,
\end{equation}
where constants $c_I$ and $d_I$ have been introduced in order to
take into account different amplitude and phase modulations for each
observable. The reduced Fisher matrix for network turns out to be
exactly the same as the reduced matrix for a single response. This
is the special case of the general property discovered by R. Prix
(\cite{Prix}, Ch. IIIC) that grid resolution in the parameter space
is independent of the number of detectors.

For low frequencies the frequency derivative $\dot{\omega}$ is small and there is no need
to include this parameter in the search. Then the linear model simplifies to
\begin{equation}
\label{smn3}
s_{3I}(t)= c_I h_0\cos{(\omega t+A\cos{\Omega t}+B\sin{\Omega
t}+\phi_0+d_I)},\;\;\;\;\;\;\;\;I=A,E,T
\end{equation}
and the corresponding reduced Fisher matrix $\tilde{\Gamma}_3$ reads
\begin{equation}\label{appFM3red}
\tilde{\Gamma}_3 =
\left(%
\begin{array}{cccc}
  \frac{1}{12}              & 0                     & -\frac{1}{2 \pi n}   \\
  0                         & \frac12               & 0   \\
 -\frac{1}{2\pi n}          & 0                     & \frac12 \\
\end{array}%
\right)  .
\end{equation}

\section*{Acknowledgments}
A.B. and A.K. would like to acknowledge hospitality
of the Max Planck Institute for Gravitational Physics in
Potsdam and Hannover, Germany where part of this work was done.
We would also like to thank Michele Vallisneri from mock LISA data challenge
Steering Committee for help in understanding the format of the challenge data files
and conventions. The work of A.B. and A.K. was supported in part by MNiSW
grant no. N N203 387237. S.B was supported  in part by DFG grant SFB/TR 7 ``Gravitational Wave Astronomy'' and
by DLR (Deutsches Zentrum f\"ur Luft- und Raumfahrt).


\begin{thebibliography}{99}
\bibitem{optimal} Note that the optimal combinations are not unique.
\end{thebibliography}
\end{document}